\documentclass[amsmath,twocolumn,twoside,11pt,a4paper]{article}

\usepackage[textwidth=16.5cm,textheight=24.5cm,headheight=0.4cm,
headsep=1.6cm,footskip=1cm,top=3cm,hmargin=2.25cm,voffset=-0.2cm]{geometry}

\pdfoutput=1

\usepackage{fancyhdr}
\usepackage{graphicx}
\usepackage{amsmath}
\usepackage{amssymb}
\usepackage{slashed}

\def\nue{\nu_e}
\def\nuebar{\bar{\nu}_e}

\def\munu{\mu_{\nu}}
\def\mub{\rm{\mu_B}}
\def\Enu{E_{\nu}}
\def\cpkkd{\rm{kg^{-1}k\eVee^{-1}day^{-1}}}
\def\rnusp{\rm{\phi ( \bar{\nu_e} ) }}
\def\s2tw{\rm{ sin ^2 \theta _W }}
\def\eVee{\rm{eV_{ee}}}

\def\dm2{\rm{\Delta m^2}}

\def\am241{\rm{ ^{241} Am }}
\def\u238{\rm{ ^{238} U }}
\def\th232{\rm{ ^{232} Th }}
\def\k40{\rm{ ^{40} K }}
\def\th232{\rm{ ^{232} Th }}
\def\u238{\rm{ ^{238} U }}
\def\cs137{\rm{^{137} Cs }}
\def\ba133{\rm{^{133} Ba }}

\begin{document}


\pagestyle{fancy}
\fancyhead{}   %
\fancyhead[LO,RE]{\small THE UNIVERSE \quad  Vol. 3, No. 4 \quad 22-37}
\fancyhead[RO,LE]{\bf\small October-December 2015}
\renewcommand{\headrulewidth}{0pt}

\fancypagestyle{plain}{%
\fancyhead{}
\lhead{\small THE UNIVERSE \quad  Vol. 3, No. 4 \quad 22-37}
\rhead{\bf \small October-December 2015}
\renewcommand{\headrulewidth}{0pt}

}

\setcounter{page}{1}


\title{
\textbf{ Taiwan EXperiment On NeutrinO $-$ \\ 
History, Status and Prospects }
}

\author{Henry Tsz-King Wong 
\footnote{email: htwong@phys.sinica.edu.tw}  \\
\footnotesize  
\textit{Institute of Physics, Academia Sinica, Taipei 11529, Taiwan.}
}

\date{}
\twocolumn[
  \begin{@twocolumnfalse}
    \maketitle
    \begin{abstract}

We present an overview of the 
foundation, evolution, contributions and future prospects
of the TEXONO Collaboration and its research programs 
on neutrino physics and dark matter searches at
the Kuo-Sheng Reactor Neutrino Laboratory in Taiwan
and, as a founding partner of the CDEX program, at the 
China Jinping Underground Laboratory in China.

    \end{abstract}
    \vspace{0.7cm}
  \end{@twocolumnfalse}
  ]

{
  \renewcommand{\thefootnote}%
    {\fnsymbol{footnote}}
    \footnotetext[1]{email: htwong@phys.sinica.edu.tw}
}


\section{Introduction}

The themes of the research programs of the
TEXONO (Taiwan EXperiment On NeutrinO) Collaboration 
are on the studies of 
low energy neutrino and dark matter physics. 
The Collaboration 
realized the first particle physics experiment
in Taiwan at the 
Kuo-Sheng Reactor Neutrino Laboratory (KSNL)
and, through the process, 
the first basic research
collaboration among researcher scientists 
from Taiwan and China~\cite{science2003}.
The efforts of the starting decade
catalyzed and laid the foundation to
the establishment of the 
China Jinping Underground Laboratory (CJPL) in China, 
together with its first generation China Dark matter
EXperiment (CDEX).
 
The history and evolution 
of the TEXONO story is given
in the following Sections.
The scientific objectives,
status, results and prospects of  
the neutrino physics program at KSNL,
as well as the underground physics program at CJPL,
are discussed.
Also surveyed is the 
theory program inspired by 
the experimental activities.

\section{History and Evolution}

\subsection{Foundation}

Phenomenal growth in basic and applied 
research has been taking place in the Asia Pacific
region in the decades of 1980's and 1990's~\cite{nature1996}. 
As the economy strengthened, 
research projects in new and advanced subjects
were initiated. 
Research infrastructures,
resources and positions were made available. 
Research directions and traditions were explored
and defined, with far-reaching consequences
beyond their original subject matters. 

Activities in experimental particle physics
started in Taiwan in the early 90's. 
The starting projects involved participation
in international experiments, including
L3, CDF, Belle, AMS, RHIC-PHOBOS, with 
contributions on various detector hardware, 
data acquisition software and data analysis projects. 

It became natural, almost inevitable, that
serious thoughts were given to an attempt to 
``perform an experiment at home'', where
local researchers would take major responsibilities 
in its conception, formulation, design, construction,
execution and scientific harvest.
Chang Chung-Yun (University of Maryland,
while on a sabbatical year at the Academia Sinica (AS))
and Lee Shih-Chang (AS) initiated
a research program towards such goals in 1996.
This ambition soon found a resonating chord with
Zheng Zhi-Peng (then Director of the Institute of
High Energy Physics (IHEP), Beijing, China) who
mobilized his Institute to participate.
Therefore, the project would carry with it 
an additional pioneering spirit of being
the first collaboration in basic
research among scientists from Taiwan and China,
standing on a decade of mutual visits and 
academic exchanges. 
It was obvious that 
many scientific, technical, administrative
and logistic issues 
have to be ironed out to make
advances on these virgin grounds.

The author (H.T. Wong) was recruited by AS
in 1997 to take up the challenges of realizing such
visions. The Chinese groups were headed
by Li Jin (senior physicist at IHEP).
The TEXONO Collaboration was formed, where
the founding partners comprised institutes
from Taiwan (AS, Institute of
Nuclear Energy Research, Taiwan Power Company 
Nuclear Station II, National Taiwan University
and National Tsing-Hua University), 
China (IHEP,
China Institute of Atomic Energy (CIAE), 
Nanjing University) and the USA (University
of Maryland).
The operation center has since been at
AS in Taiwan,
and the AS-group has been leading and coordinating
the efforts.

By the mid-2000's, the TEXONO Collaboration
has established research facilities and infra-structures,
formulated interesting research programs 
and produced world-level scientific results.
International partners 
from India (Banaras Hindu University)
and Turkey (Middle East Technical University,
Dokuz Eyl\"{u}l University) joined the
research program 
and contribute in various major items. 
In particular, an international
graduate student training scheme 
was set up as part of the operation.
Numerous graduate students from China, India
and Turkey have stationed long-term
at AS to pursue research within of the
TEXONO program and produced research theses
from the outcomes.

\subsection{Research Directions and Strategies}

Anomalous results from solar and atmospheric neutrino
measurements~\cite{pdg-nuosc} 
were gathering momentum in the 1990's,
cumulating in the presentation of evidence of
neutrino oscillation by the Super-Kamiokande
experiment in 1998~\cite{neutrino1998}.
The case of having missing energy density in the
Universe in form of Dark Matter was also getting
increasingly compelling. Non-accelerator based particle
physics was at its early stage and constituted
a good opportunity for a start-up research 
group to move into.

Reactor neutrino was identified in
the foundation days as a 
realistic platform for the TEXONO program.
One needs neutrino sources to study neutrino
physics, and reactor neutrino is an intense
and understood neutrino source, available for free 
as a by-product of commercial power
reactor operation, and allows systematic studies
and control through Reactor ON/OFF comparison.
And $-$ most importantly, there are operating
power reactor plants in Taiwan within comfortable
commuting distance from AS. 

The early target was a long baseline
reactor neutrino oscillation experiment~\cite{npbps97}. 
Feasibility studies and a liquid scintillator R\&D program
were pursued~\cite{nim-liqscin} in the first years. 
However, intense competition in the world stage 
called for the necessity to re-consider such directions.
The Chooz experiment just began to produce results
while KamLAND was already advanced in securing 
resources and has started 
hardware construction~\cite{neutrino1998}. 
It was necessary for the TEXONO program to
identify its niche, based on honest assessment 
of its strength in terms of accessible resources, 
manpower pool, experience and expertise.

The aspired goal
was that the first experiment should
on its own be able to produce valid scientific results.
The science subjects evolved to the studies
of neutrino interactions and properties, which
benefit from a location close to the reactor core
having an intense neutrino flux.
This would be complementary to
the neutrino oscillation
programs being pursued world-wide~\cite{pdg-nuosc},
where the experiments would
require baseline of kilometers or longer, 
translating directly into large detector size 
while the optimal detector technology (liquid scintillator)
and its technical details have mostly been
defined.

Reactor neutrino experiments before TEXONO
were all based on measurements of events at 
M$\eVee$ (electron-equivalent energy $\eVee$ is
used throughout this article as unit to detector
response, unless otherwise stated) 
or above, therefore only sampling the tail of
the reactor neutrino spectra.
The TEXONO program would open the previously-unexplored 
detector window in the low energy regime.
To realize these goals,
we selected detector techniques where
``the best technology is in the market''
$-$ namely, scintillating crystal detectors~\cite{ap-cryscin} 
and germanium (Ge) ionization detectors.
With the benefits of hindsights, 
this important strategic decision
allowed a new group  with ``zero background'' 
in running a particle
physics experiment to get lifted off 
and start its flight in a 
relatively short time.

The stage is thus set for the construction
of the KSNL reactor laboratory and for the formulation
of the details of its research programs.

As a record and for completeness, 
the TEXONO group, together with 
international partners, has considered
the possibility and performed feasibility
studies for a reactor measurement~\cite{theta13WP}
of the oscillation angle $\theta_{13}$ in
the early days of its formulation in 2003.
The merits with both the Kuo-Sheng Power Plant and the
new ``Nuclear Station IV'' at Lung-Men 
were investigated. 
Compared with other site proposals,
the Kuo-Sheng location lacks high mountains in
the appropriate distance, whereas Lung-Men Power Plant,
while having larger overburden, suffered from the lack
of a definite plan of starting operation 
(it {\it still} does not operate in 2016!). 
In addition, both plants are two-cores facilities,
weak in total neutrino flux compared to other 
international proposals based at locations 
with six cores or more.
Accordingly, this line of
research was not pursued after the
initial round of investigations.
A Taiwan team with three university groups
subsequently participated 
in the Daya Bay experiment which provided
the best measurement of $\theta_{13}$~\cite{dayabay}
at the six-core Daya Bay Nuclear Power complex
in southern China.

\section{Kuo-Sheng Reactor Neutrino Laboratory 
and the TEXONO Research Programs} 

\subsection{The Facility}

\begin{figure}
\begin{center}
{\bf (a)}\\
\includegraphics[width=8.0cm]{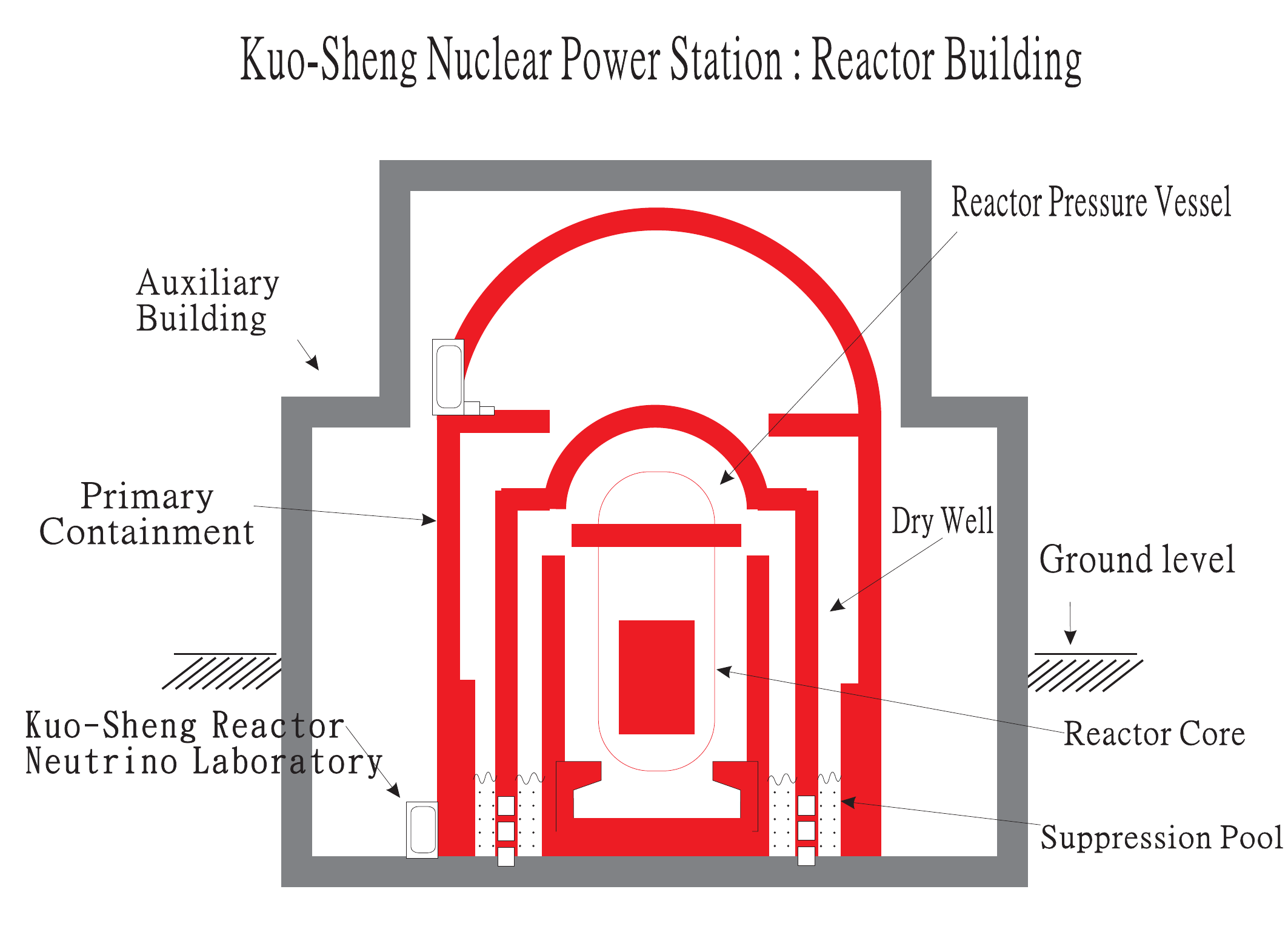}\\
{\bf (b)}\\
\includegraphics[width=8.0cm]{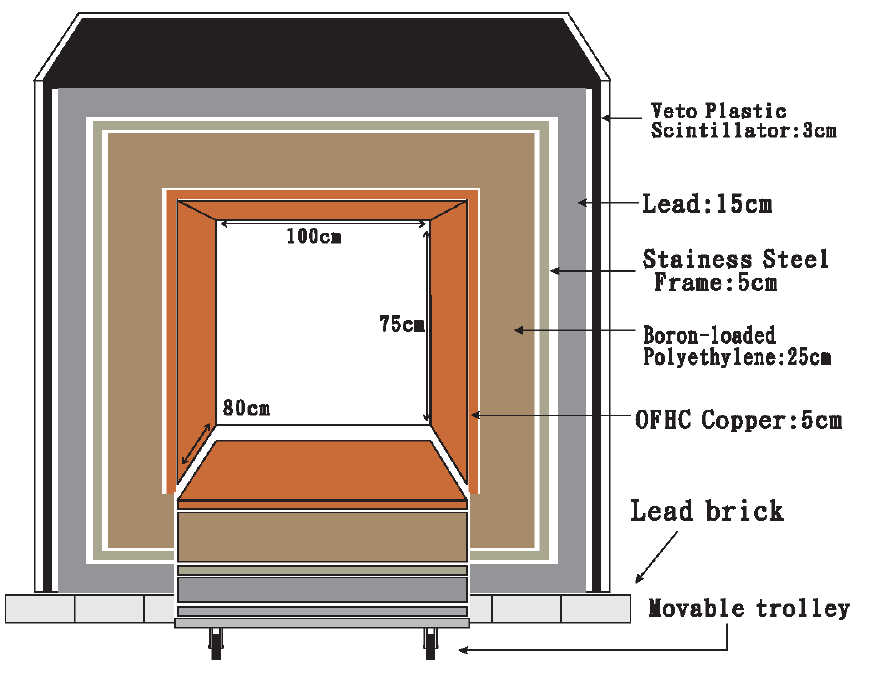}
\caption{
(a) Schematic side view, not drawn to scale,
of the Kuo-Sheng Nuclear Power Station
Reactor Building,
indicating the experimental site.
The reactor core-detector distance is about 28~m.
(b) Schematic layout of the general purpose
inner target space,
passive shieldings and cosmic-ray veto panels.
}
\label{fig::ksnlsite}
\end{center}
\end{figure}

The Kuo-Sheng Reactor Neutrino Laboratory (KSNL)
is located at a distance of 28~m from the core \#1
of the Kuo-Sheng Nuclear Power Station operated
by the Taiwan Power Company
at the northern shore of Taiwan.
The nominal thermal power output is 2.9~GW. 
Conceptual design discussions were initiated in 1997.
First physics data taking started in July 2001.
A schematic view is depicted in Figure~\ref{fig::ksnlsite}a.

A multi-purpose ``inner target'' detector space of
100~cm$\times$80~cm$\times$75~cm is
enclosed by 4$\pi$ passive shielding materials
which have a total weight of about 50 tons.
The shielding provides attenuation
to the ambient neutron and gamma background, and
consists of, from inside out,
5~cm of OFHC copper, 25~cm of boron-loaded
polyethylene, 5~cm of steel, 15~cm of lead,
and cosmic-ray veto scintillator panels.
The schematic layout of of the shielding
structure is shown in Figure~\ref{fig::ksnlsite}b.

Different detectors can be placed in the
inner space for the different scientific programs.
The detectors are read out by a general purpose
electronics and data acquisition (DAQ) systems.
Earlier versions of home-made 
electronics and VME-based DAQ~\cite{early-eledaq}
evolves into current version with a commercial-based
PXI-DAQ system with Flash Analog-to-Digital convertors
and Field Programmable Gate Array which provides 
real time processing capabilities, with 
DAQ-software via LabView packages.

The reactor laboratory is connected 
via telephone line 
(internet access not available to the reactor buildings)
to the home-base laboratory at AS, where remote access
and monitoring are performed regularly. 
The data storage
capacities are about 2~Tbytes
{\it in situ} at KSNL and
500~Tbytes at the operation
base at AS. 

\subsection{Reactor Neutrino Source}

The standard operation of the Kuo-Sheng Power Plant
includes about 18~months
of Reactor ON time at nominal power
followed by about 50~days of 
Reactor outage OFF period.
Reactor operation data on the thermal power output 
and control rod status
as functions of time and locations within the core
are provided, when necessary, to the experiment by the
Power Station. 

The $\nuebar$'s emitted in
power reactors
are predominantly
produced through $\beta$-decays of
(a) the fission products, following the
fission of the four dominant fissile isotopes:
$^{235}$U, $^{238}$U, $^{239}$Pu and $^{241}$Pu,
and
(b) $^{239}$U, following
the neutron capture on the $^{238}$U fuel:
$^{238}$U(n,$\gamma$)$^{239}$U.

\begin{figure}
\begin{center}
{\bf (a)}\\
\includegraphics[width=8.0cm]{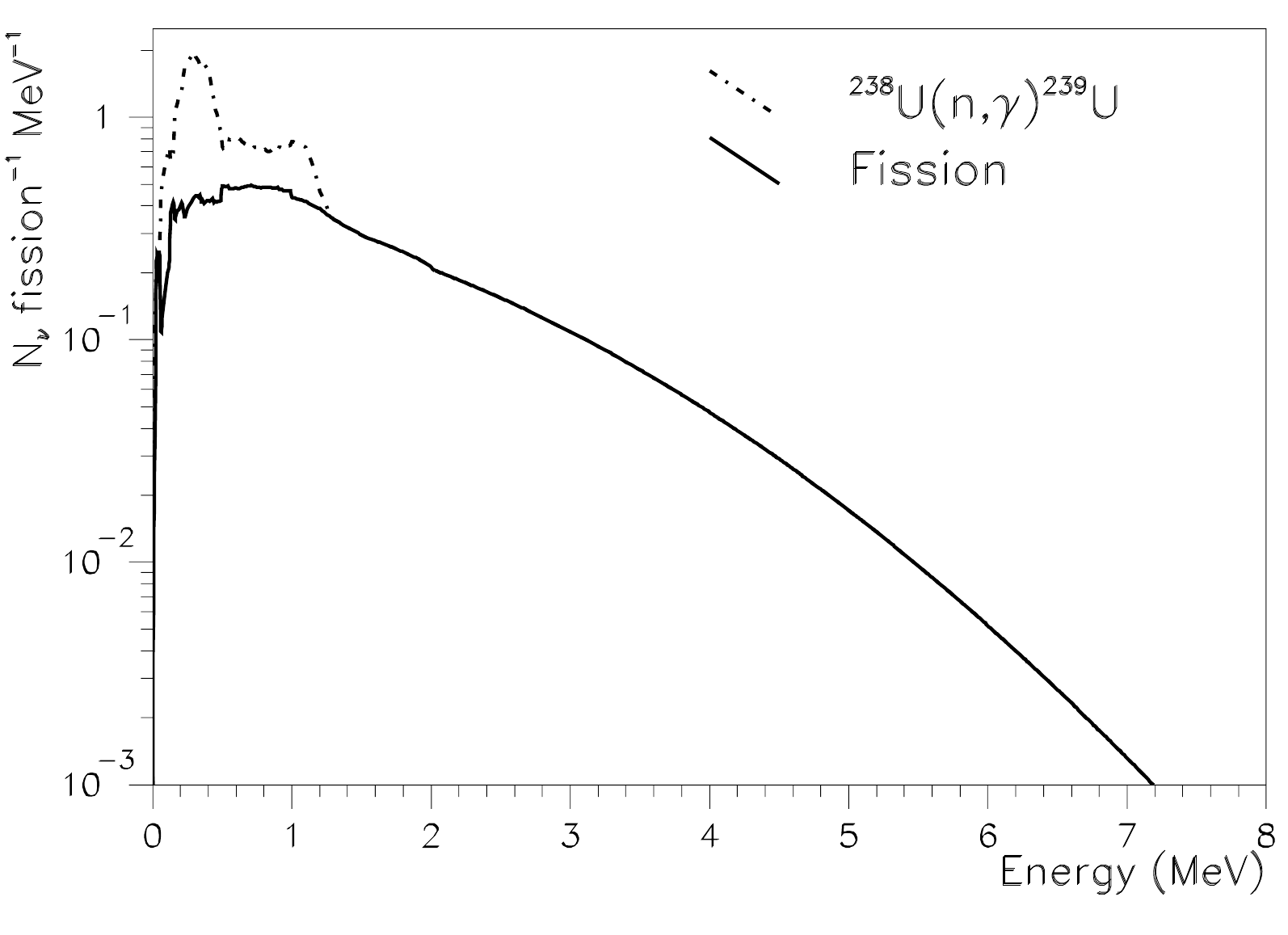}\\
{\bf (b)}\\
\includegraphics[width=8.0cm]{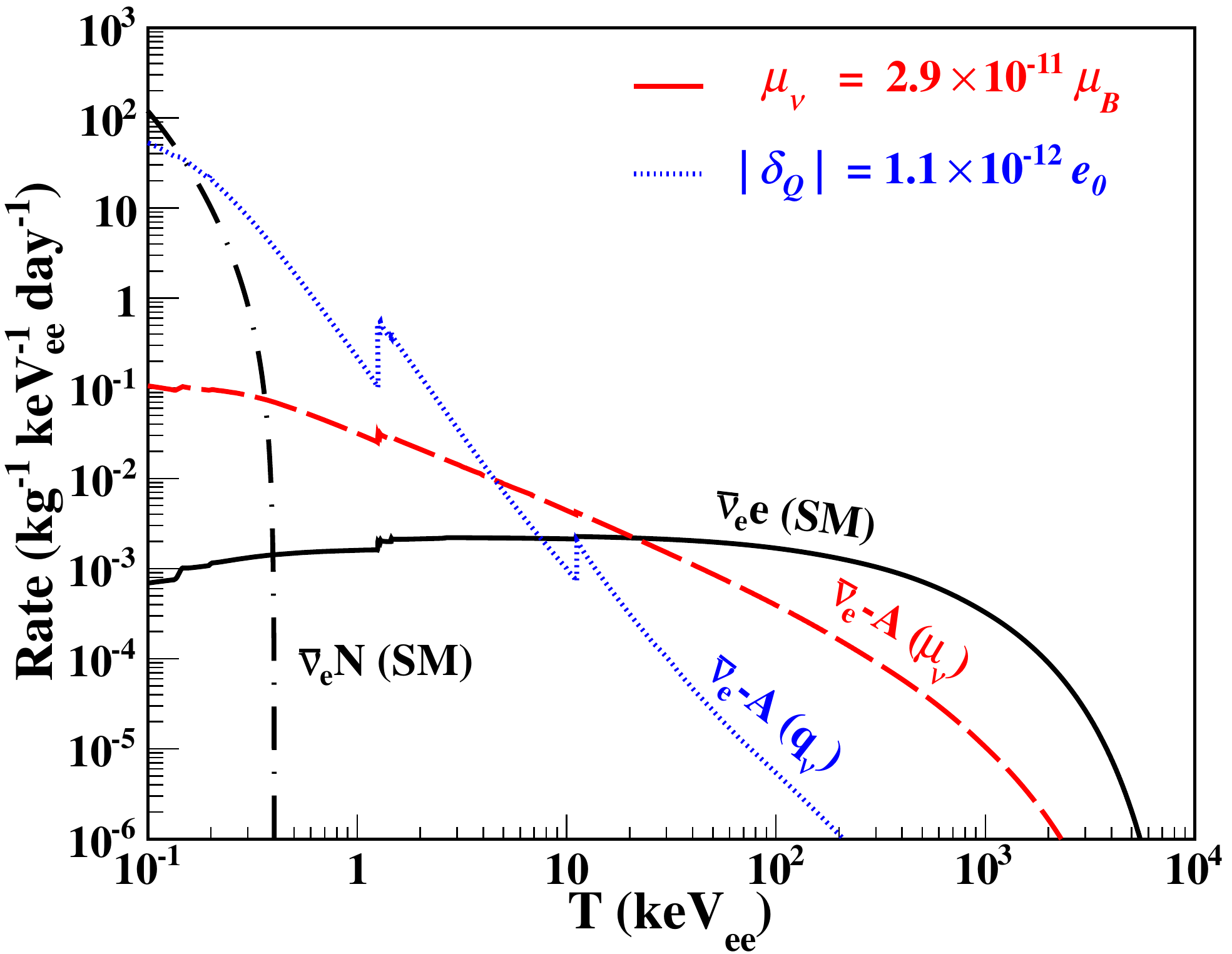}
\caption{
(a)
Typical total reactor $\nuebar$ spectrum~\cite{texonomunu,texononue}, 
normalized to per fission in a 1-MeV energy bin.
(b)
The observable recoil spectra due to 
reactor-$\nuebar$ interactions on Ge target
via Eq.~\ref{eq::nuai}
with $\phi ( \nuebar ) = 10^{13}~{\rm cm^{-2} s^{-1}}$,
neutrino magnetic moment 
and neutrino milli-charge fraction 
at the current bounds from direct experimental searches:
$\munu = 2.9 \times 10^{-11} ~ \mub$~\cite{gemma} and  
$| \delta_{\rm Q} | = 1.1 \times 10^{-12}$~\cite{numq}, 
respectively. Overlaid are the SM 
$\nuebar$-e and coherent scattering $\nuebar$-N.
Quenching effects of nuclear recoils are taken into account.
}
\label{fig::rnu+diffcs}
\end{center}
\end{figure}

The reactor neutrino spectra ($\rnusp$)
as function of neutrino energy
($\Enu$) due to the individual components 
are summed as a function of time
according to the relative contributions 
per fission~\cite{texonomunu,texononue}, 
and a typical combined reactor neutrino spectrum
is shown in Figure~\ref{fig::rnu+diffcs}a.
The typical total flux at KSNL site is
$\rm{6.4 \times 10^{12} ~ cm^{-2} s^{-1} }$.

\subsection{Low Energy Neutrino Physics}

Investigations of 
neutrino properties and interactions
can reveal physics within the Standard Model (SM)
and probe new physics beyond it (BSM).
The KSNL site provides an intense flux
of $\nuebar$ and is ideal for such investigations.
The nuclear and electron recoil differential spectra
due to reactor $\nuebar$ 
as a function of measurable recoil energy $T$
are depicted in 
Figure~\ref{fig::rnu+diffcs}b,
showing signatures due to both SM interactions 
and BSM neutrino electromagnetic effects
at the current limits.
The physics potentials becomes richer 
with lower detector threshold.

New detector technologies are necessary
to open new windows of measurable energy.
The objectives of our current 
detector R\&D program 
are to develop detectors with
modular mass of $\mathcal{O}$(1~kg), 
physics threshold 
of $\mathcal{O} ( 100 ~ \eVee )$ and background
level at threshold of $\mathcal{O}( 1~\cpkkd)$~\cite{texonoprogram}.
Intense research efforts are invested
on the operation and optimization and efficiency measurements
of Ge-detectors at sub-keV sensitivities~\cite{texono-Ge-RandD,canberra},
crucial to the studies of neutrino-nucleus coherent scattering
and to light Dark Matter searches discussed in the the
following Sections.

Complementary to and supporting the neutrino physics
and low energy detector programs is the acquisition
of low background techniques crucial for the
low count-rate experiments. Radio-purity
levels of various hardware components
were measured with different techniques.
In particular, the TEXONO-CIAE group 
explored a new arena and
performed trace contamination measurements on 
crystal and liquid scintillator materials
with accelerator mass spectrometry techniques~\cite{ciaeams}.
  
\subsubsection{Neutrino Electromagnetic Properties}

An avenue of BSM is the study
of possible neutrino electromagnetic
interactions~\cite{nuem-review}
on atomic target $A$,
via the interaction: 
\begin{equation}
\label{eq::nuai}
\nuebar ~ + ~  A ~ \rightarrow ~
\nu_X  ~ + ~ A^+ ~ + ~ e^- ~~.
\end{equation}
The target can be taken as free electrons
at $T$ above the atomic energy scale.
Otherwise, atomic physics effects
have to be taken into account~\cite{nuemai}.

The neutrino magnetic moment ($\munu$) is an 
intrinsic neutrino property
that describes possible
neutrino-photon couplings via its spin~\cite{munureview,texonomunu}.
The helicity is flipped in $\munu$-induced interactions.
Observations of $\munu$ at levels relevant to present or future
generations of experiments would strongly favor the case of 
Majorana neutrinos~\cite{naturalnessth}.
The differential cross-section  with reactor $\nuebar$
is depicted in Figure~\ref{fig::rnu+diffcs}b,
and is given by
\begin{equation}
\label{eq::mm}
( \frac{ d \sigma }{ dT } ) _{\munu}  ~ = ~
\frac{ \pi \alpha _{em} ^2 {\it \munu } ^2 }{ m_e^2 }
 [ \frac{ 1 - T/E_{\nu} }{T} ] ~ .
\end{equation}
above the atomic energy regions ($T > 10~{\rm keV}$ for Ge).
The $\munu$ contributions
are enhanced at low energy with modifications of 
the atomic binding energy effects~\cite{nuemai,nmmai,munuai10}. 


\begin{figure}
\begin{center}
\includegraphics[width=8.0cm]{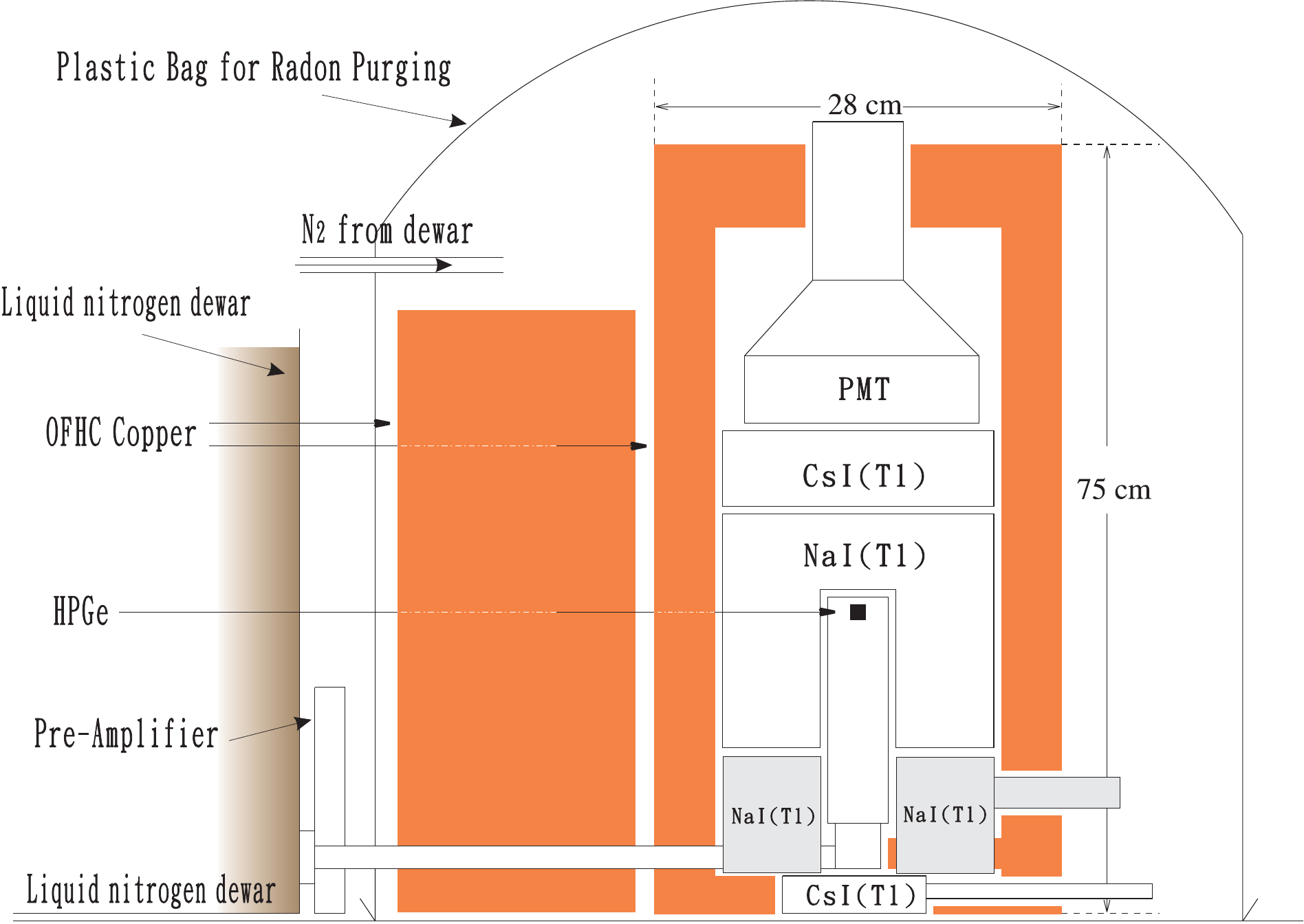}
\caption{
Schematic layout of the Ge-detector
with its anti-Compton detectors
as well as inner shieldings and
radon purge system. 
This is the baseline design
for Ge-experiments at KSNL and 
CDEX-1.
}
\label{fig::ge-blndesign}
\end{center}
\end{figure}

\begin{figure}
\begin{center}
\includegraphics[width=8.0cm]{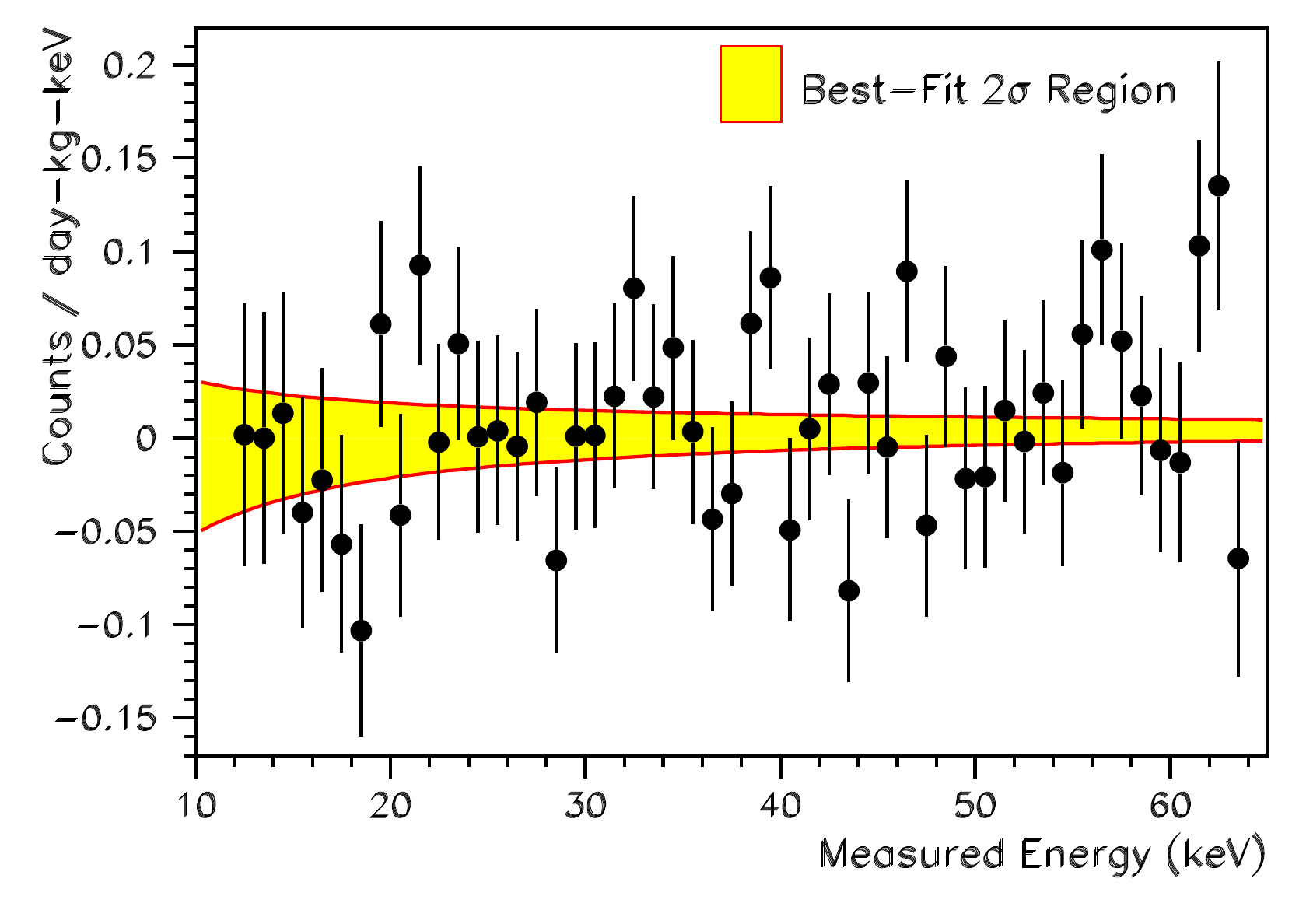}
\caption{
The residual plot 
on the combined Reactor ON data 
over the OFF-background spectra.
The allowed 2$\sigma$-band for the search
of neutrino magnetic moments is superimposed~\cite{texonomunu}.
}
\label{fig::munuresidual}
\end{center}
\end{figure}


The neutrino spin-flavor precession (SFP) mechanism, 
with or without matter
resonance effects in the solar medium,
has been used to explain solar neutrino
deficit~\cite{sfpsolar}.  This scenario
is compatible with all solar neutrino data
before 2003 until the terrestrial 
KamLAND experiment selected 
the scenario of neutrino oscillation at
large mixing angle~\cite{kamland} 
as {\it the} solution to the
solar neutrino problem~\cite{pdg-nuosc}.

The TEXONO program pioneered the studies
of neutrino physics in the low energy ($T \ll 1~{\rm M\eVee}$) 
regime~\cite{texonomunu}.
The $\munu$-experiment adopted
an ultra-low-background 
high-purity germanium
detector of mass 1.06~kg surrounded by 
NaI(Tl) and CsI(Tl) crystal scintillators
as anti-Compton detectors, as schematically
depicted in Figure~\ref{fig::ge-blndesign}. 
The setup was placed in the inner volume
within the shielding structure of Figure~\ref{fig::ksnlsite}b.
A detection threshold of 5~k$\eVee$ and 
a background level of 1~$\cpkkd$ at KSNL
near threshold were achieved.

The reactor $\rnusp$ below 
2~MeV is poorly modelled and contributed
to systematic uncertainties~\cite{lenu} to
earlier experiments at the M$\eVee$ range~\cite{munu}.
At $T \ll \Enu \sim {\rm M\eVee}$,  
the potential $\munu$-signal rates 
is much increased due to the {\it 1/T} dependence
of Eq.~\ref{eq::mm},
and is significantly higher than the
SM $\nuebar$-e ``background''
making its uncertainties less important.
The $\munu$-rate is mostly
independent of $\Enu$ at 
$T \sim$10-100~k$\eVee$, such that the $\munu$-rates
depend only on the well-known total reactor neutrino flux
but not the details of $\rnusp$.
thereby reducing the systematic uncertainties.

Based on 570.7 and 127.8 days of 
Reactor ON and OFF data, respectively,
a limit of 
\begin{equation}
\label{eq::munulimit}
\munu (\nuebar)  < 7.4 \times 10^{-11} ~ \mub
\end{equation}
at 90\% confidence level (CL)  was derived.
This result improved over existing limits~\cite{munu}
and probed the $\munu$-SFP scenario 
at the relevant range to the solar neutrino problem~\cite{sfpsolar}.
The residual Reactor ON$-$OFF spectrum
is displayed in Figure~\ref{fig::munuresidual}.

An analogous process to $\munu$-interactions
is the neutrino radiative decay~\cite{rdkmunu}
$\nu_i ~ \rightarrow ~ \nu_j ~ + ~ \gamma$
where a change of the
neutrino helicity-states takes place
and a final-state real photon is produced.
The decay rate $\Gamma _{ij}$ and
the decay lifetime $\tau _{ij}$
is related to $\mu_{ij}$ via
\begin{equation} 
\label{eq::rdk}
\Gamma_{ij} = \frac{ 1 }{ \tau _{ij} } =
\frac{1}{8 \pi} \frac{( m_i^2 - m_j^2 ) ^ 3}{m_i^3}
\mu_{ij}^2  ~ ~ ,
\end{equation}
where $m_{i,j}$ are the masses for
$\nu_{i,j}$.
The $\munu$-limit in Eq.~\ref{eq::munulimit}
translates to indirect bounds:
\begin{eqnarray}
\frac{\tau_{13}}{m_{1}^3} ( {\rm I:} \nu_1 \rightarrow \nu_3 )
& > & 3.2 \times 10 ^{27}  ~  {\rm s / eV^3} \nonumber
\\
\frac{\tau_{23}}{m_{2}^3} ( {\rm I:} \nu_2 \rightarrow \nu_3 )
& > & 1.2 \times 10 ^{27}  ~  {\rm s / eV^3} 
\\
\frac{\tau_{21}}{m_{2}^3} ( {\rm N/I:} \nu_2 \rightarrow \nu_1 )
& > & 5.0 \times 10 ^{31}  ~  {\rm s / eV^3 }  \nonumber
\end{eqnarray}
for the normal(N) or inverted(I) 
neutrino mass hierarchies~\cite{pdg-nuosc} at 90\% CL,
and are much more stringent than those from the direct
search experiments.

Experiments with similar 
baseline design of Figure~\ref{fig::ge-blndesign}
are further pursued by the GEMMA experiments at
the Kalinin Reactor in Russia, 
and the current $\munu$-limit~\cite{gemma} 
exceeds that of Eq.~\ref{eq::munulimit}.
Additional work on the KSNL Ge-data
by the TEXONO group derived the flux
and placed new constraints on the magnetic
moments of $\nu_e$~\cite{texononue2005}, 
as well as on possible axion emissions
from the reactor~\cite{texonoaxion2007}.
Another analysis studied the production
and decay of the $^{73}$Ge$^*$(1/2$^-$) metastable
states and placed constraints on neutral-current
excitation cross-sections~\cite{texonoge73}.


\begin{figure}
\begin{center}
\includegraphics[width=8.0cm]{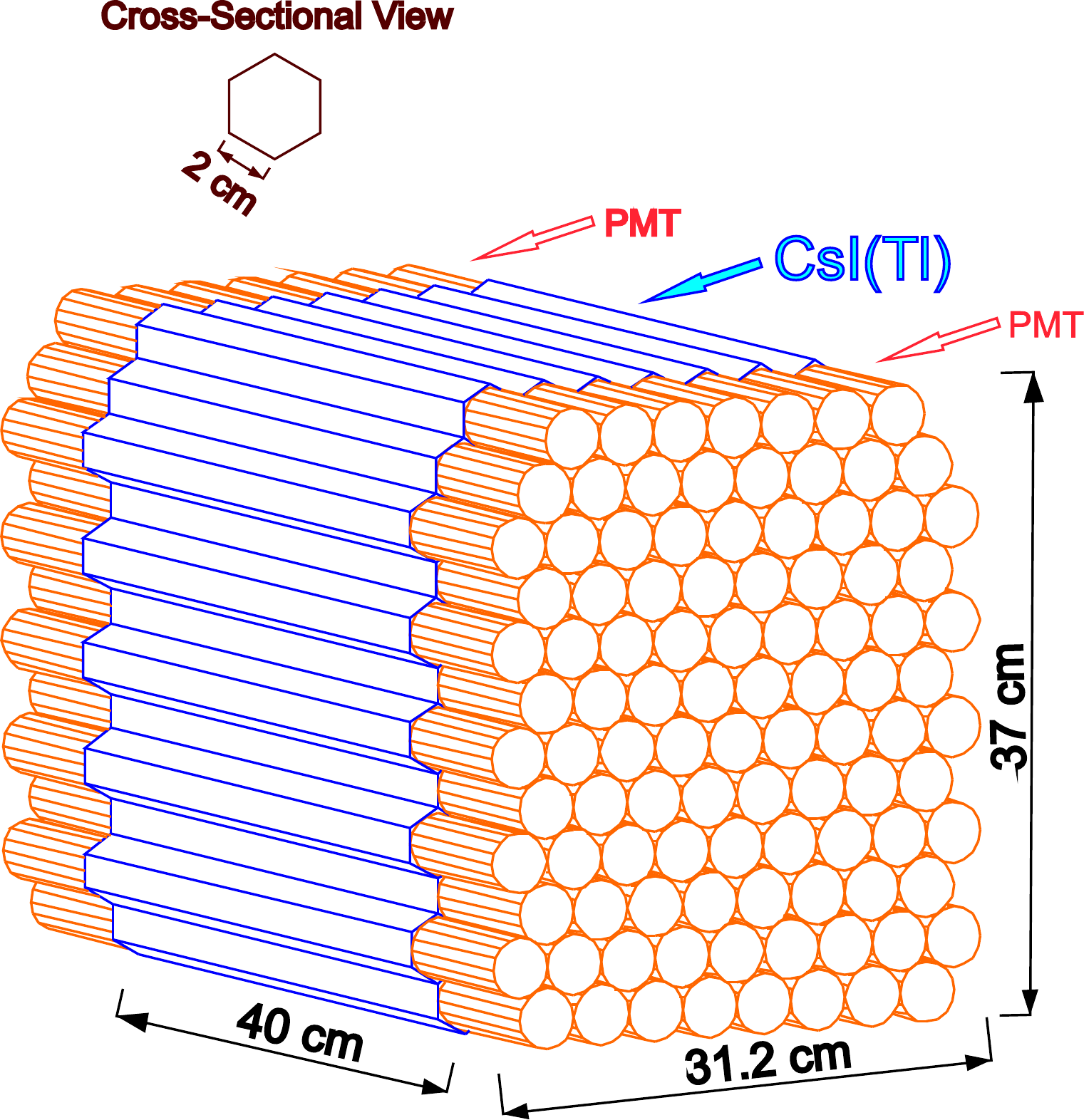}
\caption{\label{fig::csi} 
Schematic drawing of the CsI(Tl)
scintillating crystal array
for the KSNL $\nuebar$-e scattering
measurements~\cite{texononue}.
Light output is recorded
by PMTs at both ends.
}
\end{center}
\end{figure}

\begin{figure}
\begin{center}
{\bf (a)}\\
\includegraphics[width=8.0cm]{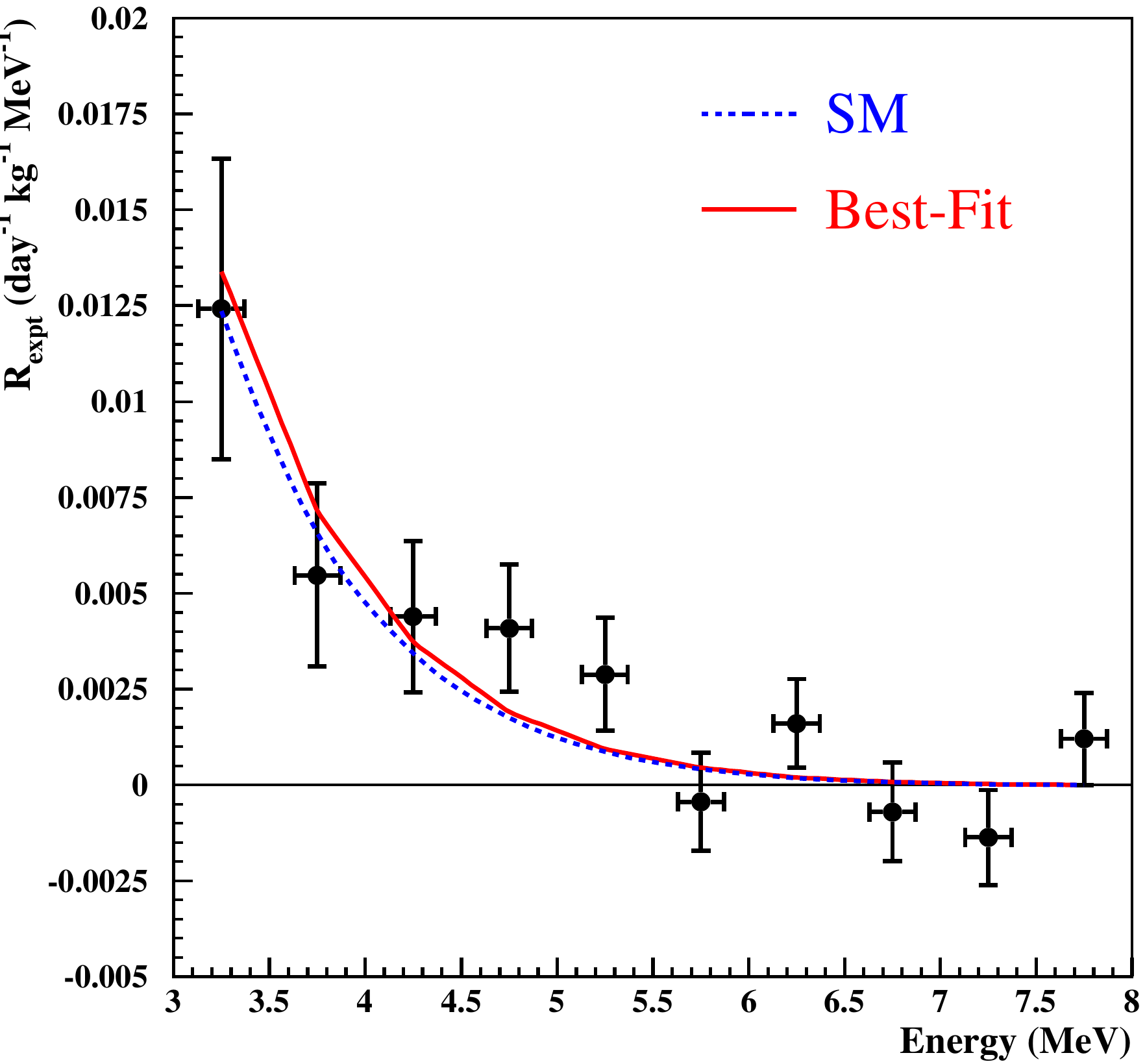}\\
{\bf (b)}\\
\includegraphics[width=8.0cm]{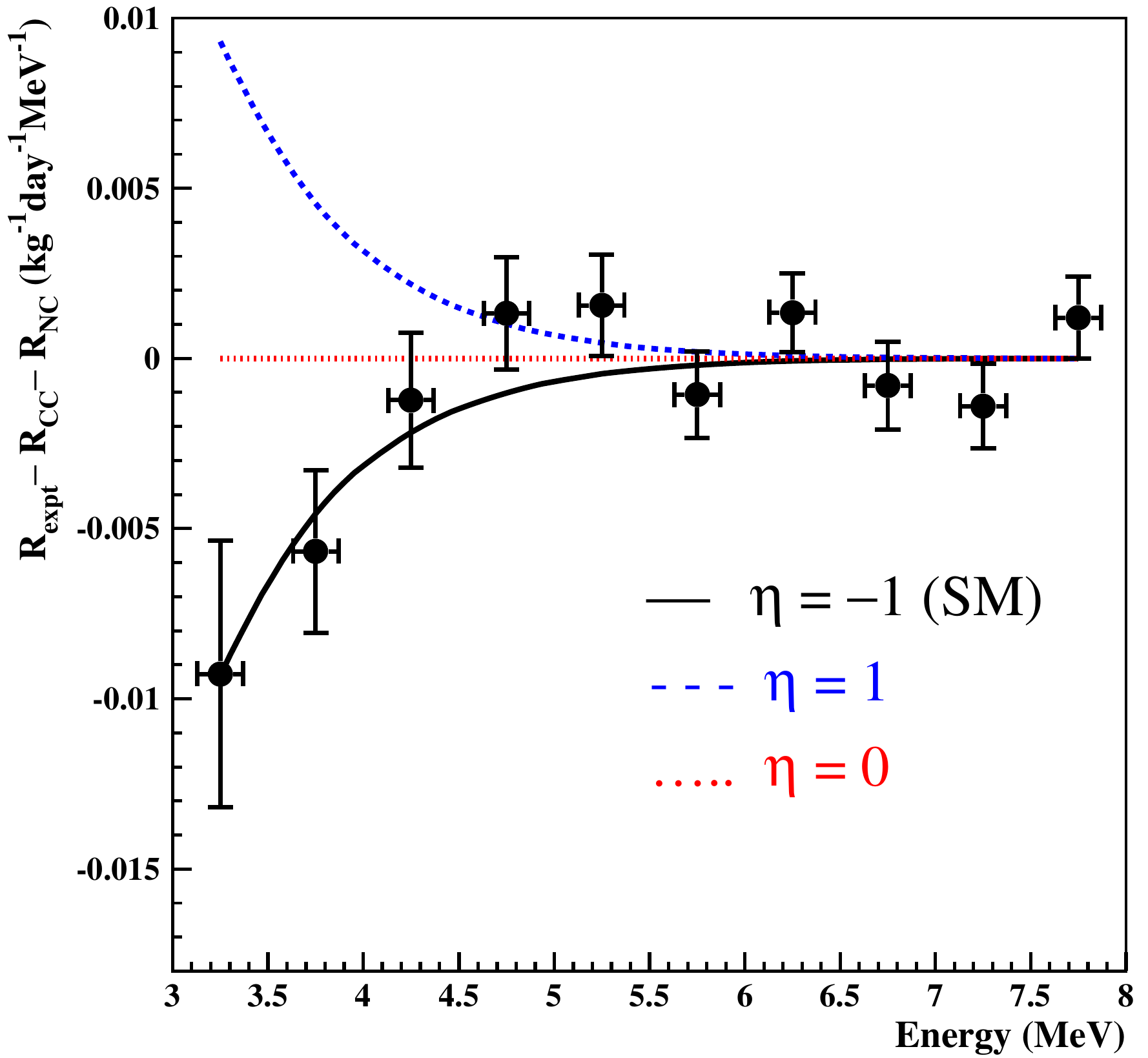}
\caption{ \label{fig::nue-spectra}
(a)
The combined 
residual spectrum of Reactor ON data 
over background~\cite{texononue}
in the 3$-$8~M$\eVee$ energy region.
The blue and red lines correspond to the SM expectations
and to the best-fit of the data, respectively,
showing excellent agreement.
(b)
The measurement of interference term based on
a residual spectrum of subtracting the charged-
and neutral-currents contributions from the data,
in the 3$-$8 M$\eVee$ energy range.
The scenario of
(no,constructive,destructive)-interference
is denoted by $\eta$=(0,1,$-$1).
The measurements verify 
the SM expectation of $\eta = - 1$.
}
\end{center}
\end{figure}


\subsubsection{Neutrino-Electron Elastic Scattering}

The TEXONO program has provided the
best cross-section measurement on
 two of the fundamental leptons in Nature $-$
$\nuebar$ with electrons~\cite{texononue}: 
\begin{equation}
\label{eq::nuew}
\nuebar ~ + ~  e^- ~ \rightarrow ~
\nuebar ~ + ~  e^- ~~.
\end{equation}

Reactor $\nuebar$ provides a unique
laboratory to measure
neutrino-electron scattering,
and therefore probe electro-weak physics~\cite{pdg-electroweak}
at the  MeV momentum transfer range.
The $\nuebar$-e interaction,
together with the analogous $\nu_e$-e studied 
with accelerator neutrinos~\cite{lsndnue}, 
are among the few
processes which proceed via charged- and
neutral-currents {\it and} their interference channel. 
The SM cross-section can be written as:
\begin{eqnarray}
\left[ \frac{d\sigma}{dT}(\bar{\nu}_{e}e ) \right] _{SM} & = & 
\frac{G_{F}^{2}m_{e}}{2\pi }  \cdot  
[ ~ \left(g_{V}-g_{A}\right) ^{2}  \\
& + & \left( g_{V}+g_{A}+2\right) ^{2}\left(1-
\frac{T}{E_{\nu }}\right) ^{2}  \nonumber  \\
& - & (g_{V}-g_{A})(g_{V}+g_{A}
+2)\frac{m_{e}T}
{E_{\nu}^{2}}  ~ ] .   \nonumber
\label{eq::gvga}
\end{eqnarray}
The SM assignments to the electroweak coupling constants
are:
\begin{equation}
g_{V}=-\frac{1}{2}+2\sin ^{2}\theta _{W}\text{ \ \ \ \ and \ \ \ \ }
g_{A}=-\frac{1}{2}\label{eq_gvga} ~~~ ,
\end{equation}
where $\s2tw$ is the weak mixing angle.


A scintillating CsI(Tl) crystal detector 
array~\cite{texononue,csiRandD},
as depicted in Figure~\ref{fig::csi},
was constructed for this measurement.
The detector is as a proton-free target,
with modules packed into a matrix array, having minimal
inactive dead space due to the teflon wrapping sheets.
A total of $12 \times 9$ array was deployed
giving a total mass of 187~kg.
Each crystal module is 40~cm in length with
light output read out by photo-multipliers (PMT)
at both ends. The sum of the normalized PMT signals
provides the energy, while their difference
defines the longitudinal location. Therefore,
event reconstruction in three-dimension 
is achieved.  The fiducial volume was defined 
to be the inner crystals
with a separation of $>$4~cm 
from the PMTs at both ends.

Reactor ON/OFF data, 
of 29882/7369 kg-days strength were taken.
The Reactor ON over background residual spectrum,
as depicted in Figure~\ref{fig::nue-spectra}a
shows excellent consistency with SM predictions.
The ratio of experimental to SM cross-sections of 
\begin{equation}
\frac{R_{expt} ( \nu )}{R_{SM} ( \nu )} 
= 1.08 \pm 0.21 (stat) \pm 0.16 (sys)  
\end{equation}
was measured. 
After accounting for the charged- and neutral-current
components, the SM destructive interference 
in $\nuebar$-e interactions was verified, 
as illustrated in Figure~\ref{fig::nue-spectra}b.


\begin{figure}
\begin{center}
\includegraphics[width=8.0cm]{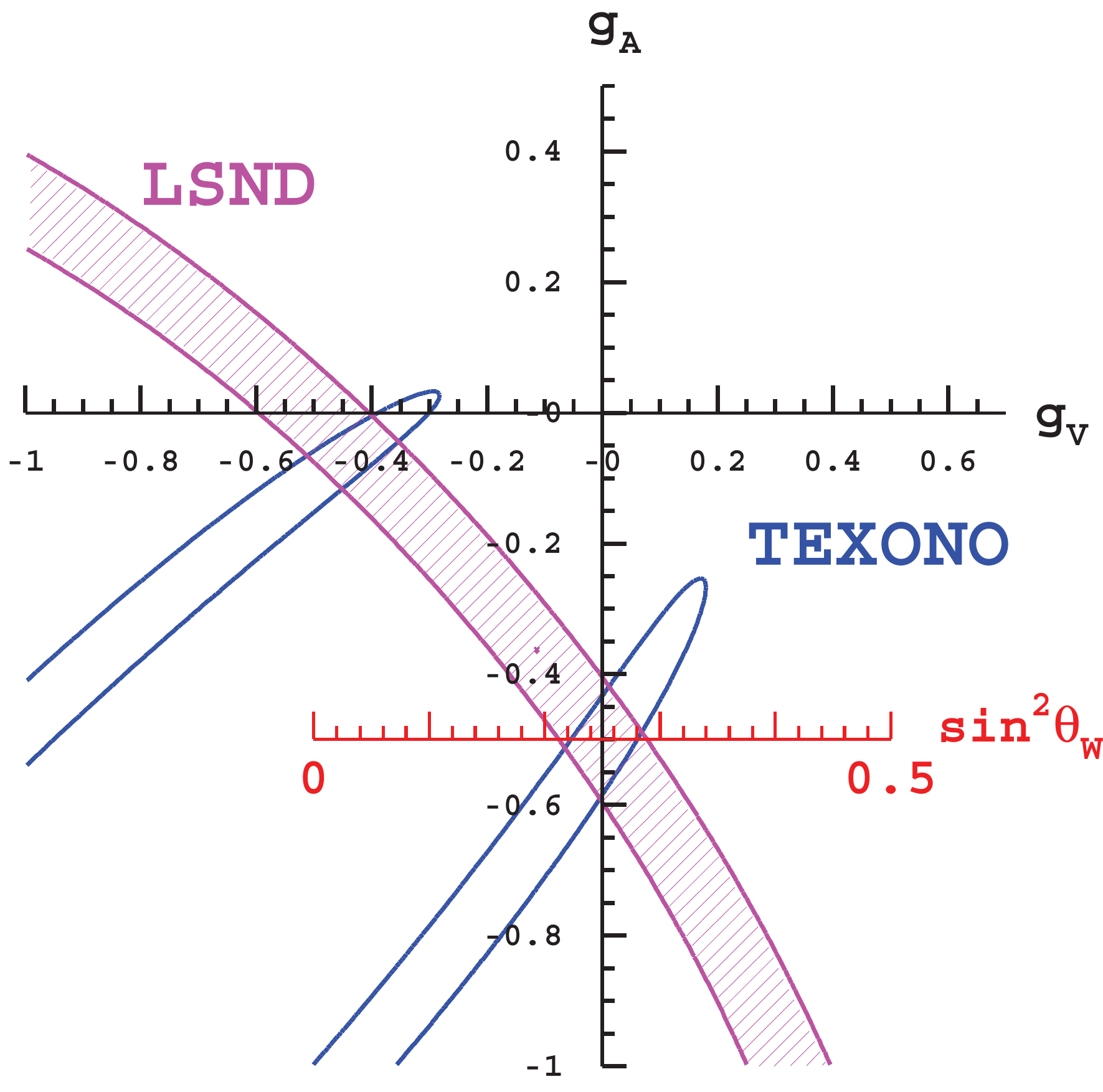}
\caption{\label{fig::gvga}
Best-fit results 
in $( g_{V} , g_{A} )$ space and in the $\s2tw$ axis
from the TEXONO-CsI(Tl) experiment at KSNL
on $\nuebar -$e~\cite{texononue} and 
the LSND experiment on $\nue -$e~\cite{lsndnue}.
The allowed regions are defined by
their corresponding statistical uncertainties.
}
\end{center}
\end{figure}


Constraints on the electroweak parameters
$( g_V , g_A )$ were placed, as
illustrated in Figure~\ref{fig::gvga}.
The corresponding weak mixing angle 
at the squared 4-momentum transfer range of
$\rm{Q^2 \sim 3 \times 10^{-6} ~ GeV^2}$
was measured:  
\begin{equation}
 \s2tw = 0.251 \pm 0.031 ({\it stat}) \pm 0.024 ({\it sys})  ~~.
\end{equation}

The consistency of the data with SM 
can be translated to bounds  on the
neutrino charge radius of
\begin{equation}
-2.1 \times 10^{-32} ~{\rm cm^{2}} 
~ < ~  \langle r_{\bar{\nu}_e}^2  \rangle ~ < ~
3.3 \times 10^{-32} ~{\rm cm^{2}} 
\end{equation}
at 90\% CL, improving over earlier limits.

\subsubsection{Neutrino-Nucleus Elastic Scattering}

The current theme of the neutrino physics program
at KSNL is on the observation of 
the elastic scattering between a neutrino and a 
nucleus ($\nu N$)~\cite{texonoprogram,nuNcoh} :
\begin{equation}
\label{eq::nuN}
\nu ~ + ~ N ~ \rightarrow ~
\nu ~ + ~ N
\end{equation}

It is a fundamental 
SM-predicted 
neutrino interaction which has never been observed.
It probes coherence effects in electroweak interactions~\cite{nuNalpha}, 
and provides a sensitive test to physics beyond SM.
The coherent interaction plays an important role 
in astrophysical processes
and constitutes the irreducible background channel to
forthcoming generation of dark matter experiments.

The maximum nuclear recoil energy for a Ge target (A=72.6) 
due to reactor $\nuebar$ is about 2~${\rm keV_{\rm nr}}$.
The quenching factor (ratio of ionization to total deposited energy), 
is about 0.2 for Ge in the 
$< {\rm 10~ keV_{\rm nr}}$ region~\cite{texono-Ge-RandD}. 
Accordingly, the maximum measurable energy 
for nuclear recoil events in Ge due to reactor
$\nuebar$ is about 300~$\eVee$. 
The typical differential spectrum 
is given in Figures~\ref{fig::rnu+diffcs}b. 
At benchmark sensitivities, 
the expected rate is of $\mathcal{O}{\rm ( 10~kg^{-1}day^{-1} )}$ 
with a signal-to-background ratio $>$50.

Improvement of the lower reach of detector sensitivity
without compromising background
is therefore crucial for such experiments,
and are the focuses of our current research.

\subsection{Dark Matter Searches}

The goal of measuring the 
neutrino-nucleus coherent scattering
process of Eq.~\ref{eq::nuN} at KSNL 
leads to the development of low threshold
Ge-detector with sub-keV sensitivity~\cite{texono-Ge-RandD}.
This detector technology
naturally brings the Collaboration
to venture into the important arena on 
direct Dark Matter searches~\cite{texonoprogram}.

Weakly Interacting Mass Particles (WIMPs, denoted by $\chi$) 
are leading candidates to the Dark Matter Problem in the
Universe~\cite{cdmpdg14}.
The elastic recoils between WIMPs 
and the nuclei
\begin{equation}
\label{eq::chiN}
\chi ~ + ~ N ~ \rightarrow ~
\chi ~ + ~ N 
\end{equation}
are the favored channel in direct dark matter
search experiments.
Consistency with observations on 
cosmological structure formation
requires that WIMPs
should be massive and their motions are non-relativistic.
The measurable nuclear recoil energy is therefore small,
such that the experimental requirements are similar to those
for $\nu N$ where low detector threshold is crucial.

\begin{figure}
\begin{center}
{\bf (a)}\\
\includegraphics[width=8.0cm]{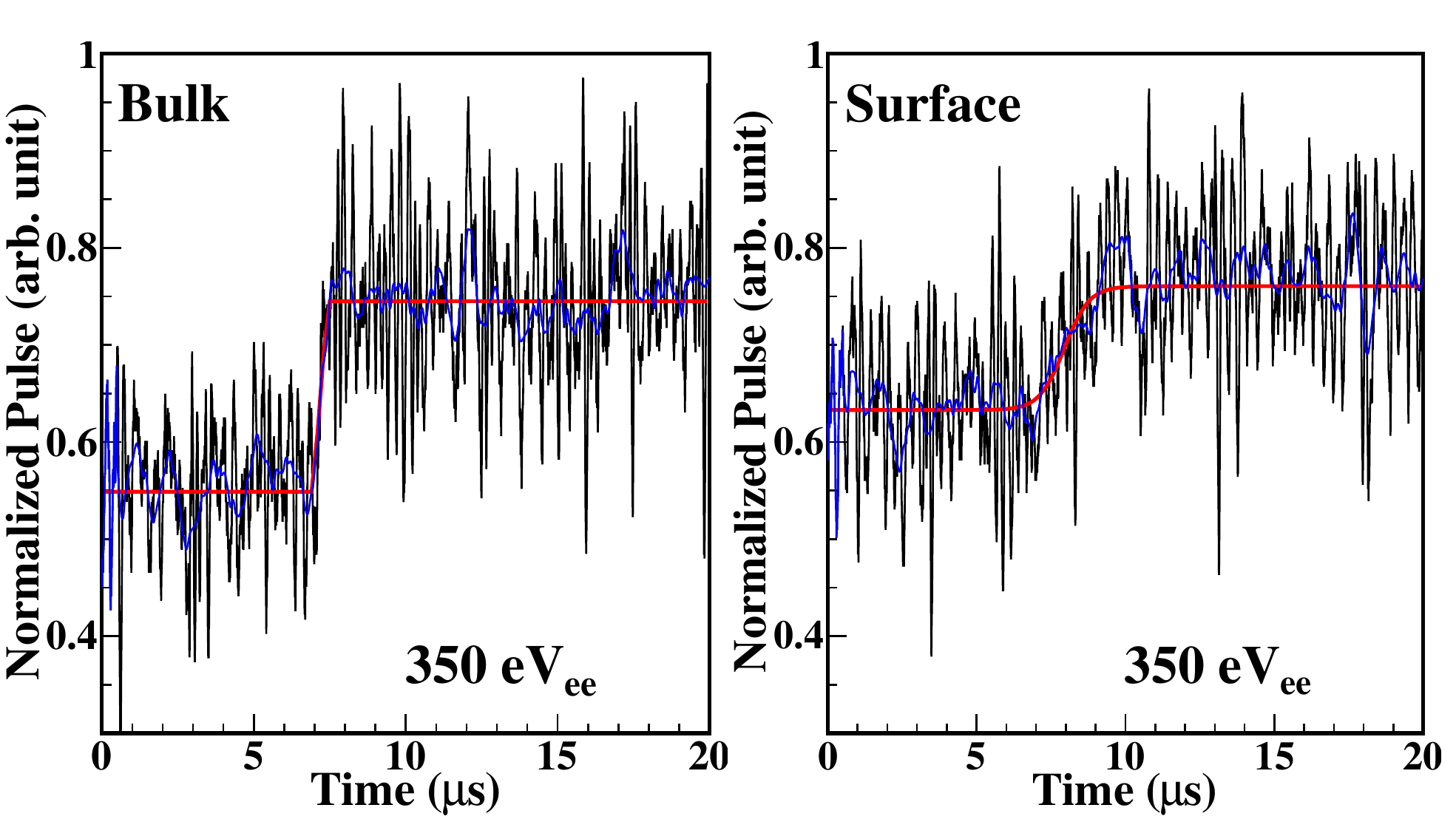}\\
{\bf (b)}\\
\includegraphics[width=8.0cm]{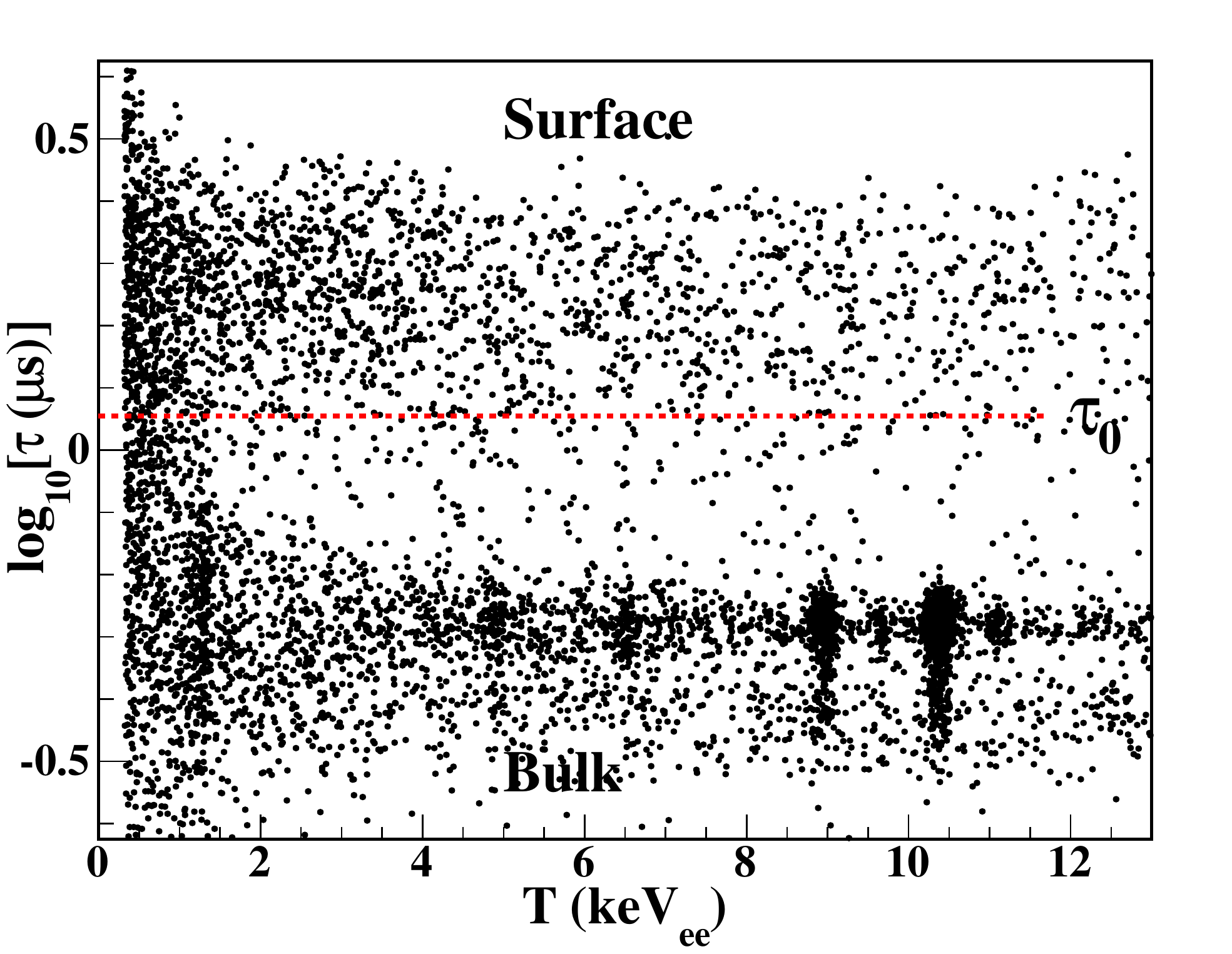}\\
{\bf (c)}\\
\includegraphics[width=8.0cm]{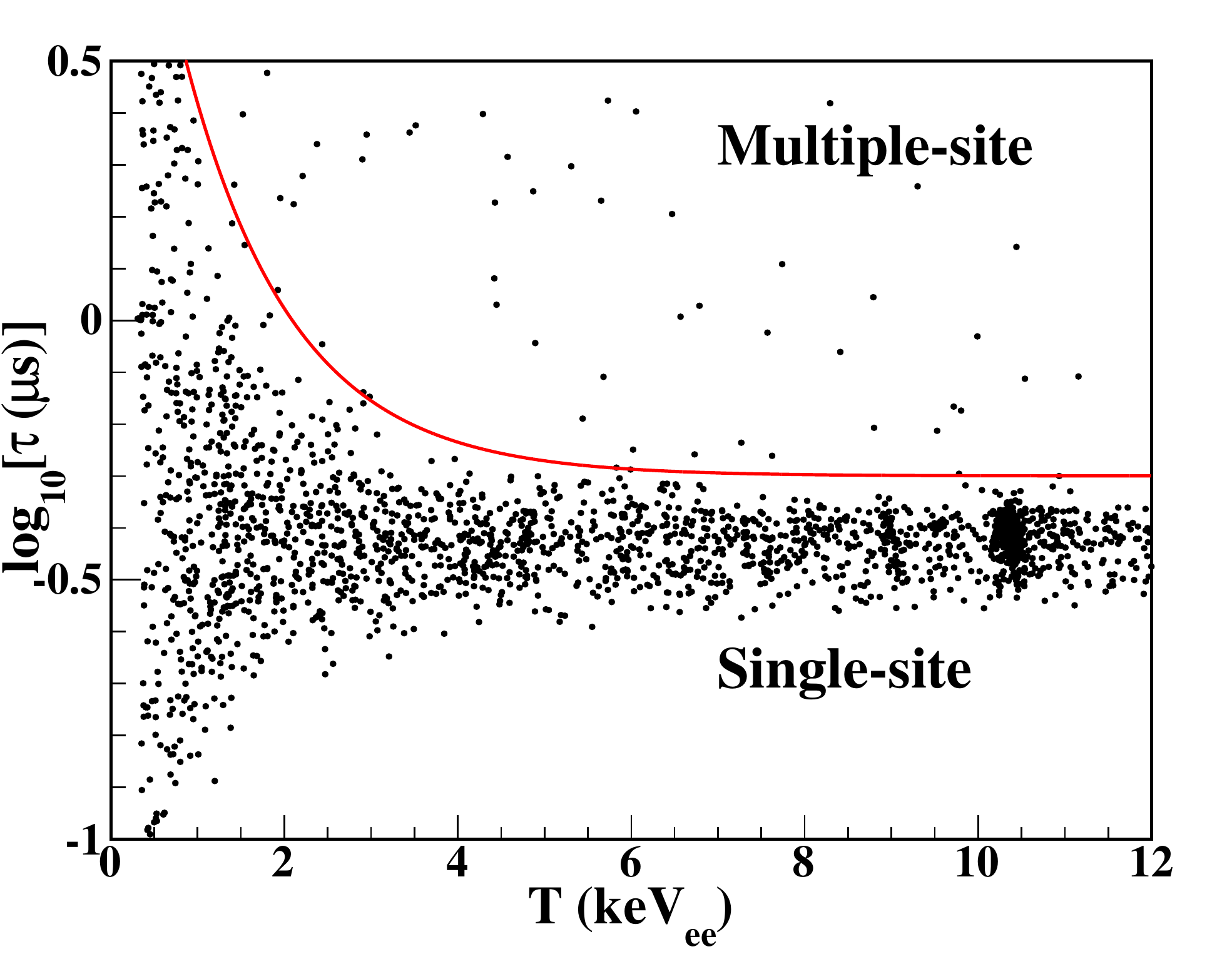}
\caption{\label{fig::tau-PCGe}
(a)
Measured pulse profiles for typical Bulk and Surface 
events~\cite{texono-Ge-RandD,bsel2014}
in p-PCGe.
(b)
The rise-time($\tau$) distribution in p-PCGe.
The selection criteria of signal events in the Bulk
is defined by the cut at $\tau_0$.
(c)
The rise-time($\tau$) distribution in n-PCGe.
There are no anomalous surface events.
}
\end{center}
\end{figure}

\begin{figure}
\begin{center}
{\bf (a)}\\
\includegraphics[width=8.0cm]{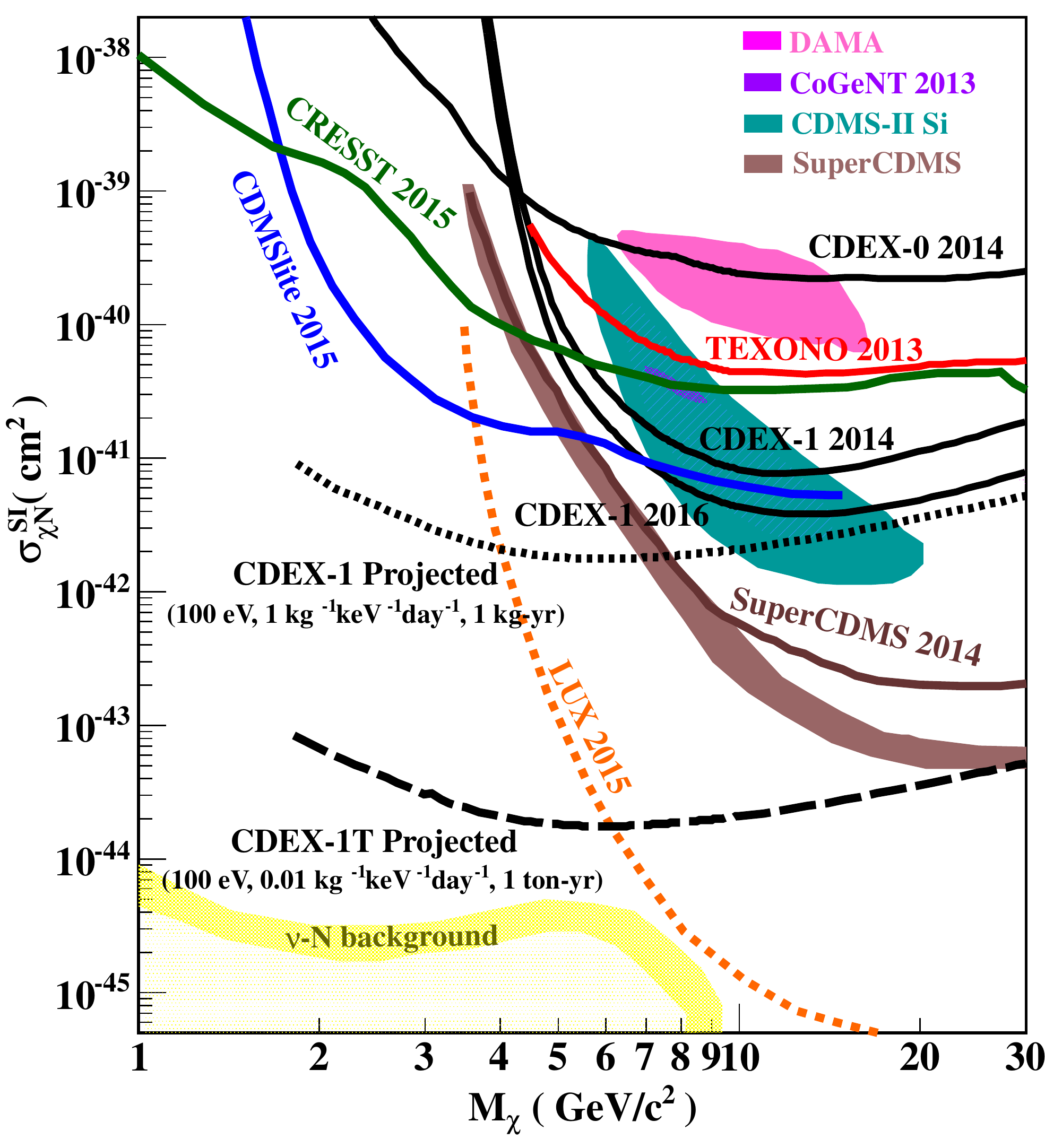}\\
{\bf (b)}\\
\includegraphics[width=8.0cm]{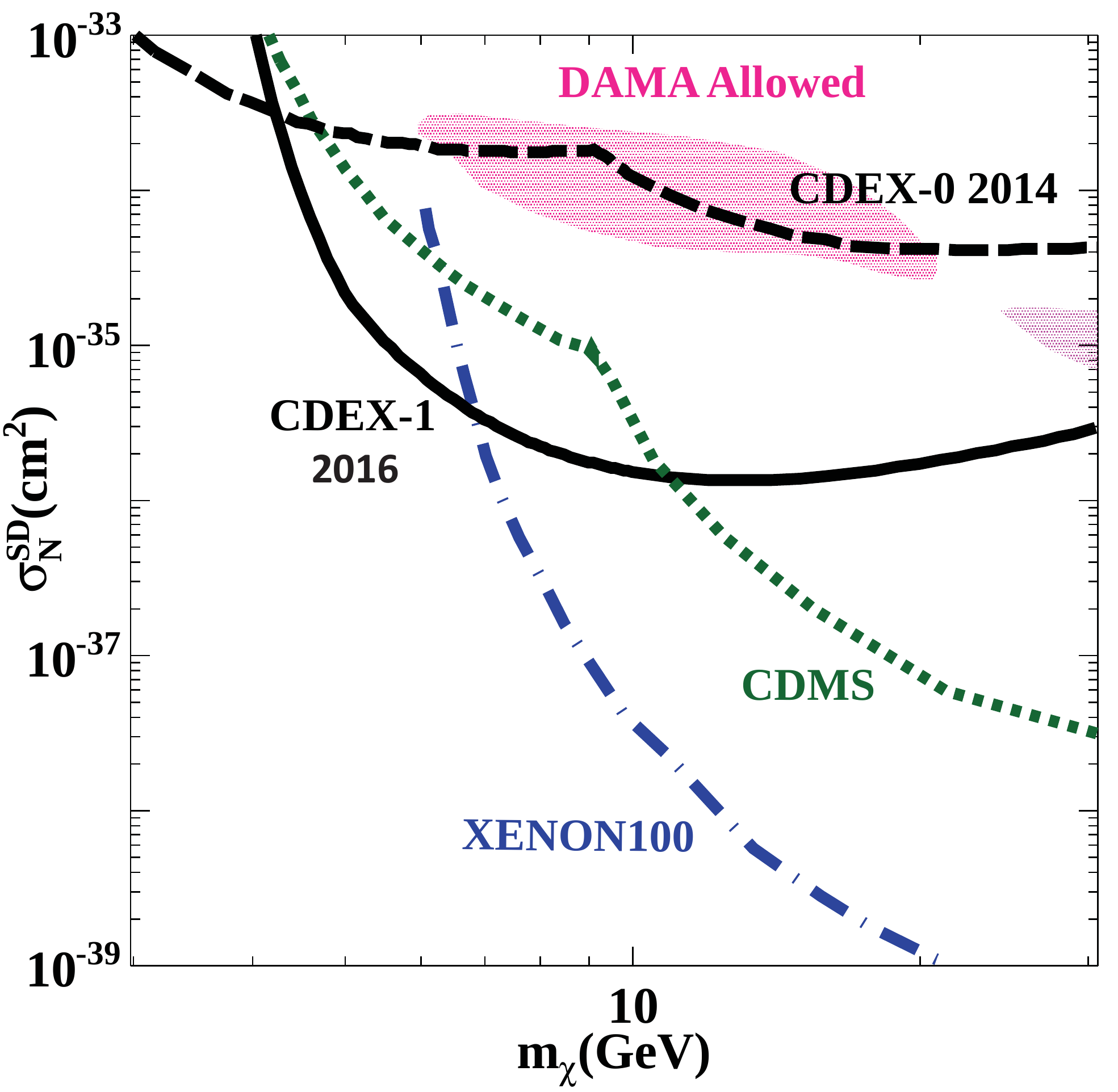}
\caption{\label{fig::explot}
Exclusion plots of $\chi N$ 
(a) spin-independent and (b) spin-dependent
interactions, showing the TEXONO~\cite{texonocdm2013}
and CDEX~\cite{cdex,cdex0} results 
together with the allowed regions and limits
from other benchmark experiments~\cite{cdmpdg14}.
}
\end{center}
\end{figure}

We opened the sub-keV detector window with pilot 
``Ultra-Low-Energy'' Ge detectors (ULEGe)~\cite{texonoprogram}
with modular mass of the order of 10~g.
Data taken with a 20~g ULEGe array at
an analysis threshold
of 220~$\eVee$ at KSNL~\cite{texonocdm2009}, 
a surface laboratory
hardly appropriate for such experiments,
already allowed the probing of new 
parameter space in the ``Light WIMPs'' region
of several GeV in mass.  

Our early efforts on WIMP searches
inspired advances
in point-contact germanium detectors (PCGe)
by a US group based on an earlier design~\cite{pcge},
realizing sub-keV sensitivity with a modular mass
at kg-scale.
The CoGeNT experiment subsequently reported
possible allowed regions for light WIMPs~\cite{cogent}
turning this into a domain of intense interest. 
The TEXONO group performed a
measurement at KSNL
with a PCGe with p-type germanium (p-PCGe) 
of 840~g fiducial mass~\cite{texonocdm2013},
following the baseline setup of Figure~\ref{fig::ge-blndesign}.

Crucial to this study is 
the bulk-surface events differentiation
at the sub-keV range~\cite{texono-Ge-RandD,bsel2014}.
As illustrated with Figure~\ref{fig::tau-PCGe}a,
the surface background events in p-PCGe detectors, 
exhibit slower rise-time and
partial charge collection compared to
bulk events which are candidates for $\chi N$-signals.
The measured rise-time distribution  
is depicted in Figure~\ref{fig::tau-PCGe}b.
In the contrary, n-type PCGe does not
have anomalous surface layer and shows
uniform rise-time, as shown in 
Figure~\ref{fig::tau-PCGe}c.

Selection schemes of bulk signal
events in p-PCGe were devised, and the
corresponding signal-retaining and background-rejecting
efficiencies were measured~\cite{bsel2014}.
Our results indicated deficiencies of the previous
approaches, and the excess events
interpreted to be WIMP candidate signals
are due to incomplete correction of the
leakage of surface background into the
bulk signal samples.
The exclusion plot 
of $\chi N$ spin-independent 
cross-section versus WIMP-mass
is displayed in Figure~\ref{fig::explot}a.

Light WIMP searches started as
an exploratory by-product of the TEXONO 
program at KSNL on sub-keV Ge-detectors
and neutrino-nucleus coherent scattering.
As we shall discuss in
Section~\ref{sect::cjpl}, these
efforts would inspire and catalyze
the realization of the deepest and
largest underground scientific facility 
at CJPL.

\subsection{Theory Programs}

The TEXONO experimental program and
the unique data at KSNL triggered 
several theory research  directions,
and establishes 
fruitful collaborations among
the experimental and theory researchers.

One direction, spearheaded by the TEXONO Turkish groups,
is to study the constraints from neutrino-electron
scattering to BSM models, especially those where
data at low-$T$, such as those from KSNL,
would provide enhanced sensitivities. 
These include~\cite{texonoBSMconstraints} 
generic non-standard interactions,
unparticle physics, non-commutative physics and
dark photon physics.

Another line is rooted in
our current experimental theme and uniqueness
of using novel germanium detectors with
sub-keV sensitivities and very good energy resolution
to explore neutrino and dark matter physics.
This technique excels in probing experimental 
signatures at the ``atomic'' energy range 
and with possible spectral structures
due to atomic ionization.
A pilot investigation~\cite{munuai10} 
set the stage on the subsequent studies
of ``atomic ionization'' cross-sections.
Theorists in Taiwan (J.W. Chen and C.P. Liu and their groups),
through collaboration with the TEXONO group,
introduced state-of-the-art theoretical tools 
in atomic physics
(MCRRPA$-$ Multi-Configuration 
Relativistic Random-Phase Approximation~\cite{mcrrpa}) 
leading to a series of results
on experimental signatures due to neutrino
electromagnetic effects~\cite{nuemai,numq,munusterilenu},
illustrated in the differential cross-sections
of Figure~\ref{fig::rnu+diffcs}b.

Some of the results provide positive feedback
to the experiment programs and data interpretation,
examples of which include:
\begin{enumerate}
\item
Studies of the ``neutrino milli-charge'' 
probe possible helicity conserving QED-like interactions.
Finiteness of the neutrino charge fraction 
($\delta _Q$) would imply 
neutrinos are Dirac particles.
It was demonstrated that
atomic ionization effects due to $\delta _Q$
lead to big enhancement
in cross-sections~\cite{numq},
as depicted in Figure~\ref{fig::rnu+diffcs}b. 
The known ratios of peaks at discrete binding 
energies provide
smoking gun signatures for positive observations.
\item
A massive sterile neutrino can have transition-$\munu$
and interact with matter to become a light SM neutrino.
If it is non-relativistic, as in the case 
of a dark matter candidate, the interaction would have 
a cross-section pole and enhancement at half its 
mass~\cite{munusterilenu}. Constraints were
derived from KSNL data. 
\item
The quantum mechanical coherence effects
of electroweak interaction 
can be quantitatively studied,
using the $\nu N$ scattering of Eq.~\ref{eq::nuN}.
We derived how the degree of coherence would
vary with realistic neutrino sources, target
materials and detection threshold~\cite{nuNalpha},
showing how the forthcoming experimental projects
can complement each other.
\end{enumerate}


\section{China Jinping Underground Laboratory
and the CDEX Research Programs}
\label{sect::cjpl}

\subsection{Foundation}

The potentials of dark matter experiments
were immediately realized after 
initial sub-keV measurements achieved with
germanium detectors~\cite{texonoprogram}.
While the main thrust of the Collaboration
remains on the development of the detector
techniques and reactor neutrino physics at KSNL
where dark matter physics is a by-product,
the TEXONO-Tsinghua University (THU) group explored
the means to turn it into dedicated experiments.

An underground site is therefore mandatory.
The THU group spearheaded a pilot project of
installing a 5~g ULEGe detector 
at the Yangyang Underground Laboratory in
Korea in 2004, supported by the KIMS group as host
of that Facility.

A construction road tunnel was completed
in 2008 under the Jinping mountains in
Sichuan province in China 
to facilitate the construction of the numerous
hydro-electric power facilities in that region
$-$ the flagship project is the Jinping-I dam
which, at 305~m, is the tallest dam in the world.
The physics communities in China immediately
recognized the opportunities and potentials. 
By 2010, agreement was made between the site owner
Yalong River Hydropower Development Company
and THU to jointly develop
and operate an underground laboratory facility − 
the China Jinping Underground Laboratory (CJPL)~\cite{cjpl-birth}. 
Civil engineering proceeded in full swing.

The TEXONO-THU group got significant
boost in its manpower and resource pool
to match the expanding engineering
demands and scientific program. 
The group evolved and emerged to form and lead a new 
CDEX (China Dark matter EXperiment)
Collaboration focusing on,
in its start-up phase, 
dark matter experiments at CJPL.
The CDEX program is led by 
Kang Ke-Jun (a former Vice President of THU).
The TEXONO Collaboration became
a founding partner and participant
of this endeavour.

\begin{figure}
\begin{center}
{\bf (a)}\\
\includegraphics[width=8.0cm]{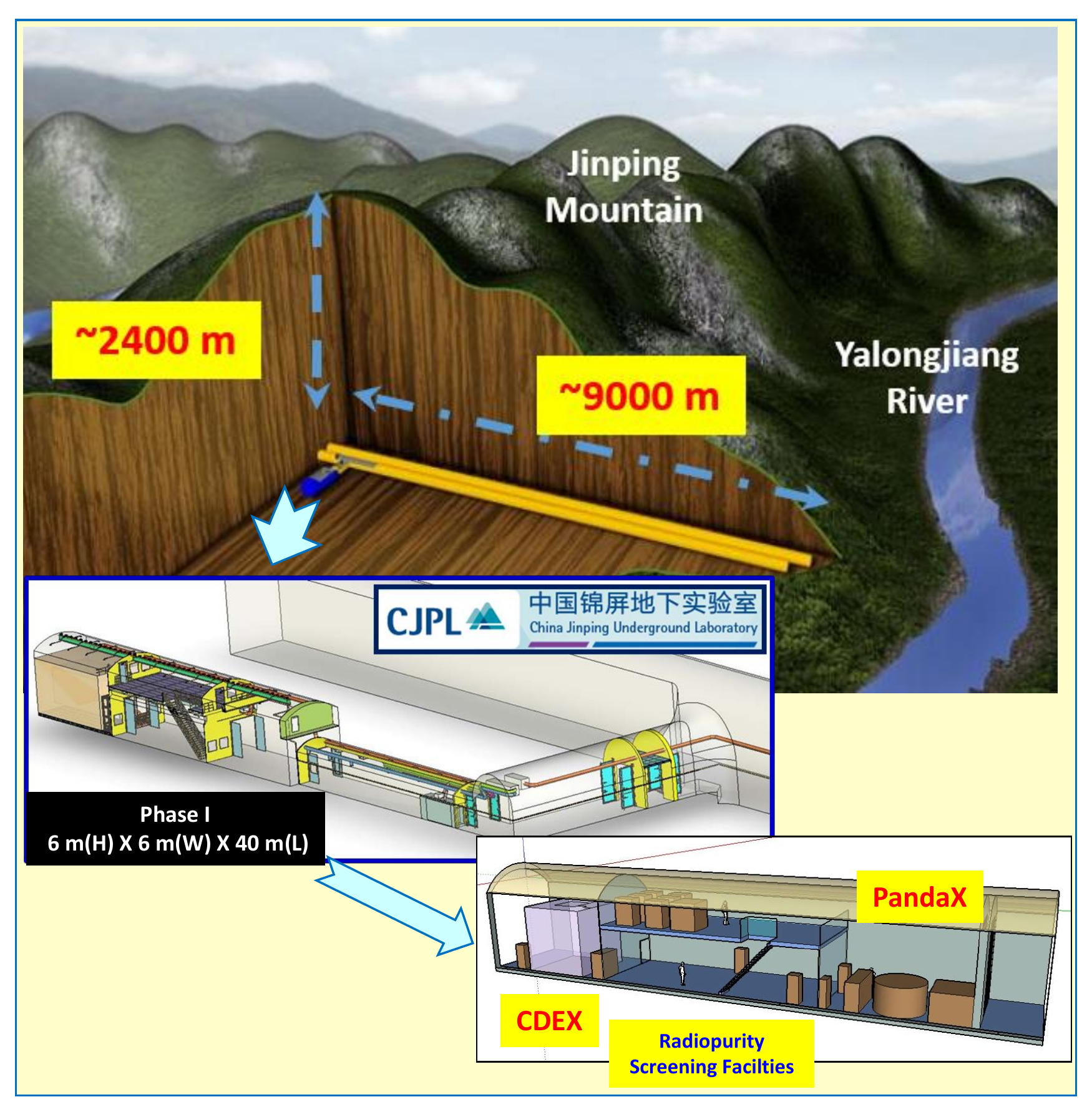}\\
{\bf (b)}\\
\includegraphics[width=8.0cm]{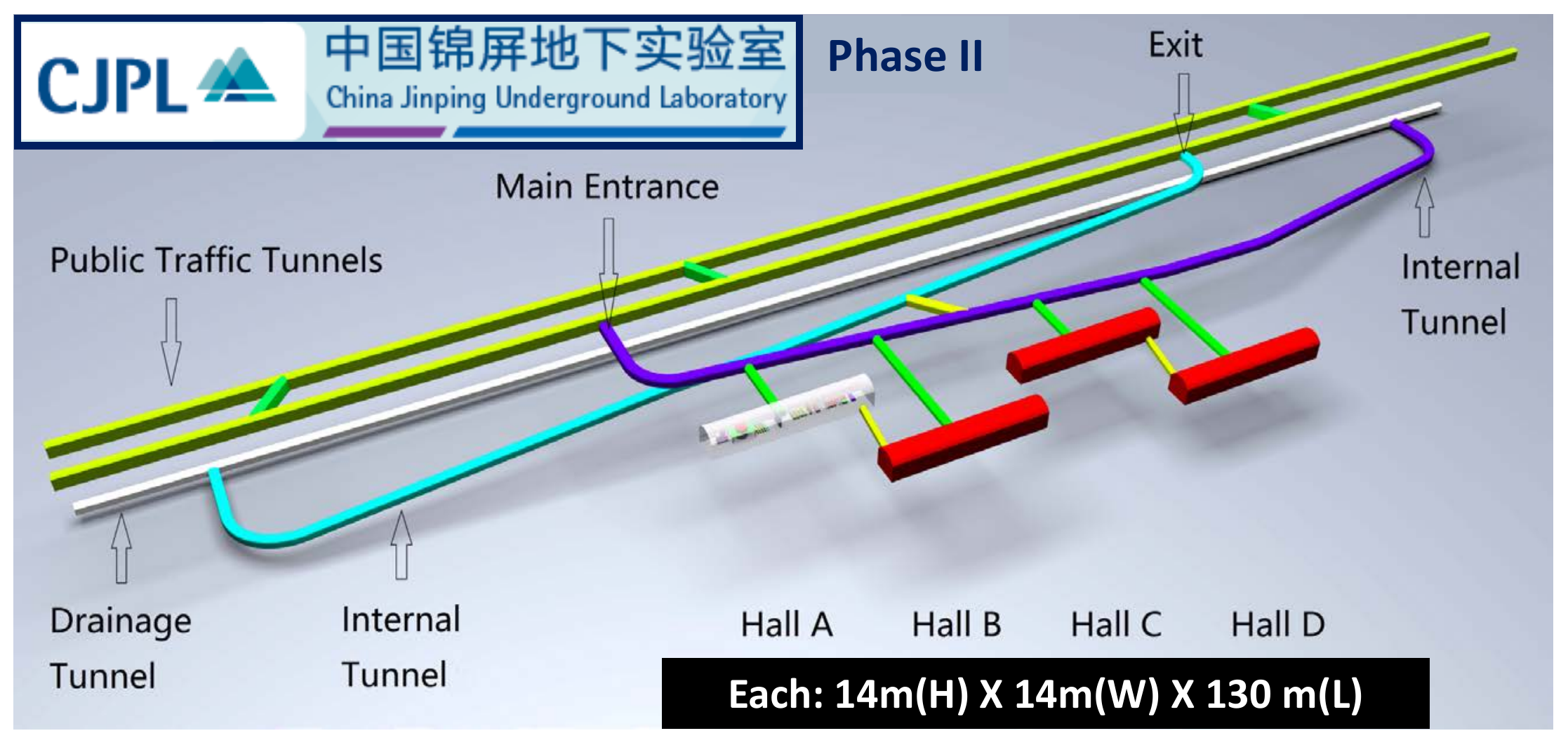}
\caption{\label{fig::cjpl} 
Schematic diagram of 
(a) CJPL~Phase-I inaugurated in 2012, 
showing the space allocation
to the CDEX and PandaX Dark Matter experiments, as well as to
the radio-purity screening facilities,
(b) CJPL~Phase-II scheduled to complete
by early 2017, showing the four experimental halls
and the tunnel systems.
}
\end{center}
\end{figure}


\subsection{The Facility}

The Facility CJPL~\cite{cjpl}
is located in Sichuan, China, and
was inaugurated in December 2012. 
With a rock overburden of about 2400~meter,
it is the deepest operating underground laboratory in
the world. 
The muon flux is measured to be
$( 2.0 \pm 0.4 ) \times 10^{-10} \rm{cm^{-2} s^{-1}}$~\cite{cjpl-cosmic}, 
suppressed from the sea-level flux by a factor of $10^{-8}$.
The drive-in tunnel access can greatly facilitate the
deployment of big experiments and large teams.
Supporting infrastructures of catering and accommodation, 
as well as office and workshop spaces, 
already exist.
All these merits make CJPL an ideal location
to perform low count-rate experiments.

As depicted schematically in Figure~\ref{fig::cjpl}a,
the completed CJPL~Phase-I consist of a laboratory hall
of dimension 6~m(W)$\times$ 6~m(H)$\times$40~m(L).
This space is currently shared by the CDEX~\cite{cdex} 
and PandaX\cite{pandax} dark matter
experiments, as well as a general purpose low radio-purity
screening facility.

Additional laboratory space for CJPL~Phase-II~\cite{cjpl2-media}, 
located about 500~m from the Phase-I site,
is currently under construction.
It will consist of four experiment halls each with dimension
14~m(W)$\times$14~m(H)$\times$130~m(L), connected by
an array of access tunnels.
The schematic layout of CJPL-II is displayed 
in Figure~\ref{fig::cjpl}b.
Upon the scheduled completion by early 2017,
CJPL will be, in addition, the largest underground
laboratory by volume in the world.

\subsection{CDEX Dark Matter Program}

The scientific theme of 
CDEX program~\cite{cdex} 
is to pursue studies of light WIMPs with p-PCGe.
It is one of the two founding experimental programs at CJPL.

\begin{figure}
\begin{center}
{\bf (a)}\\
\includegraphics[width=8.0cm]{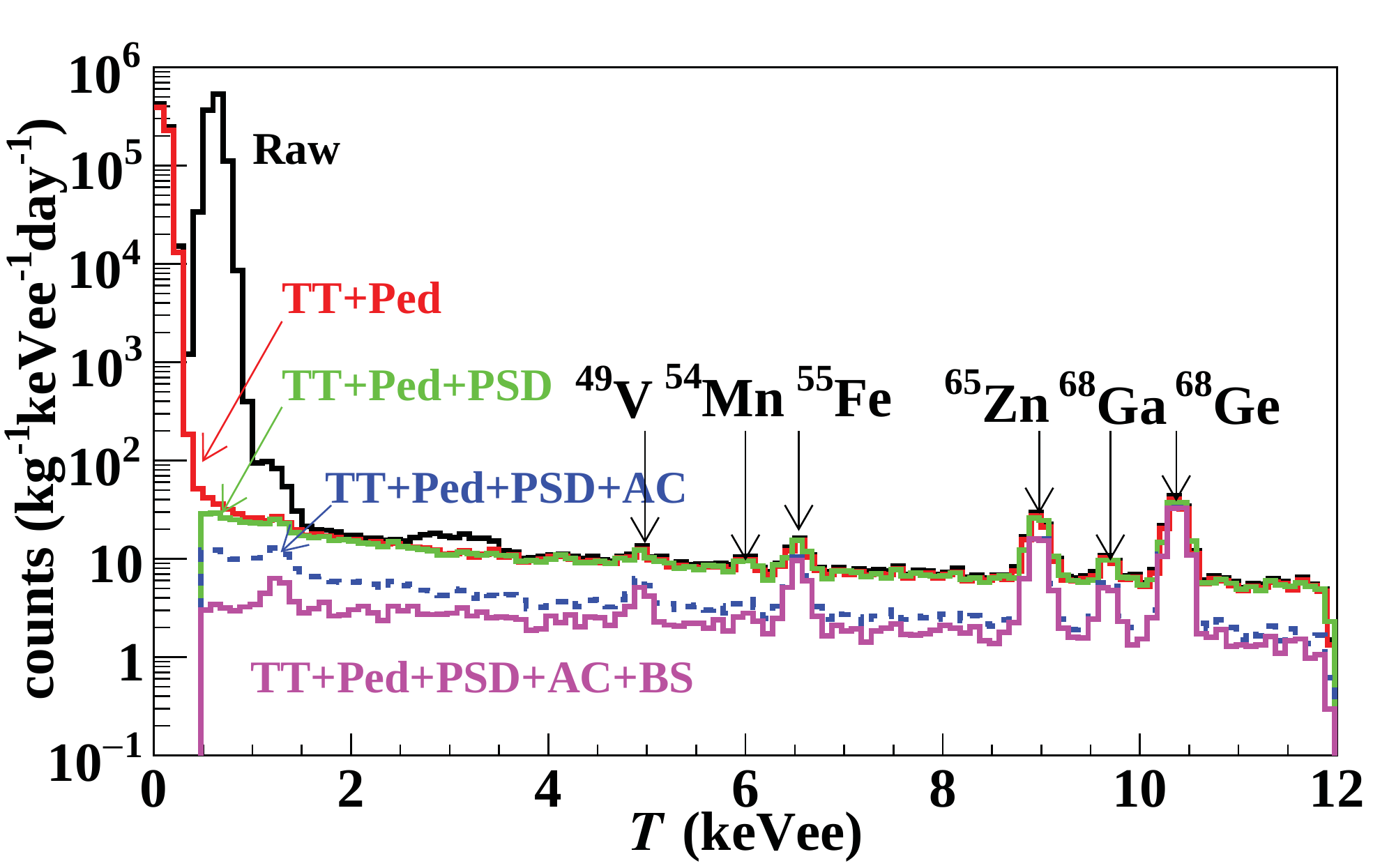}\\
{\bf (b)}\\
\includegraphics[width=7.0cm]{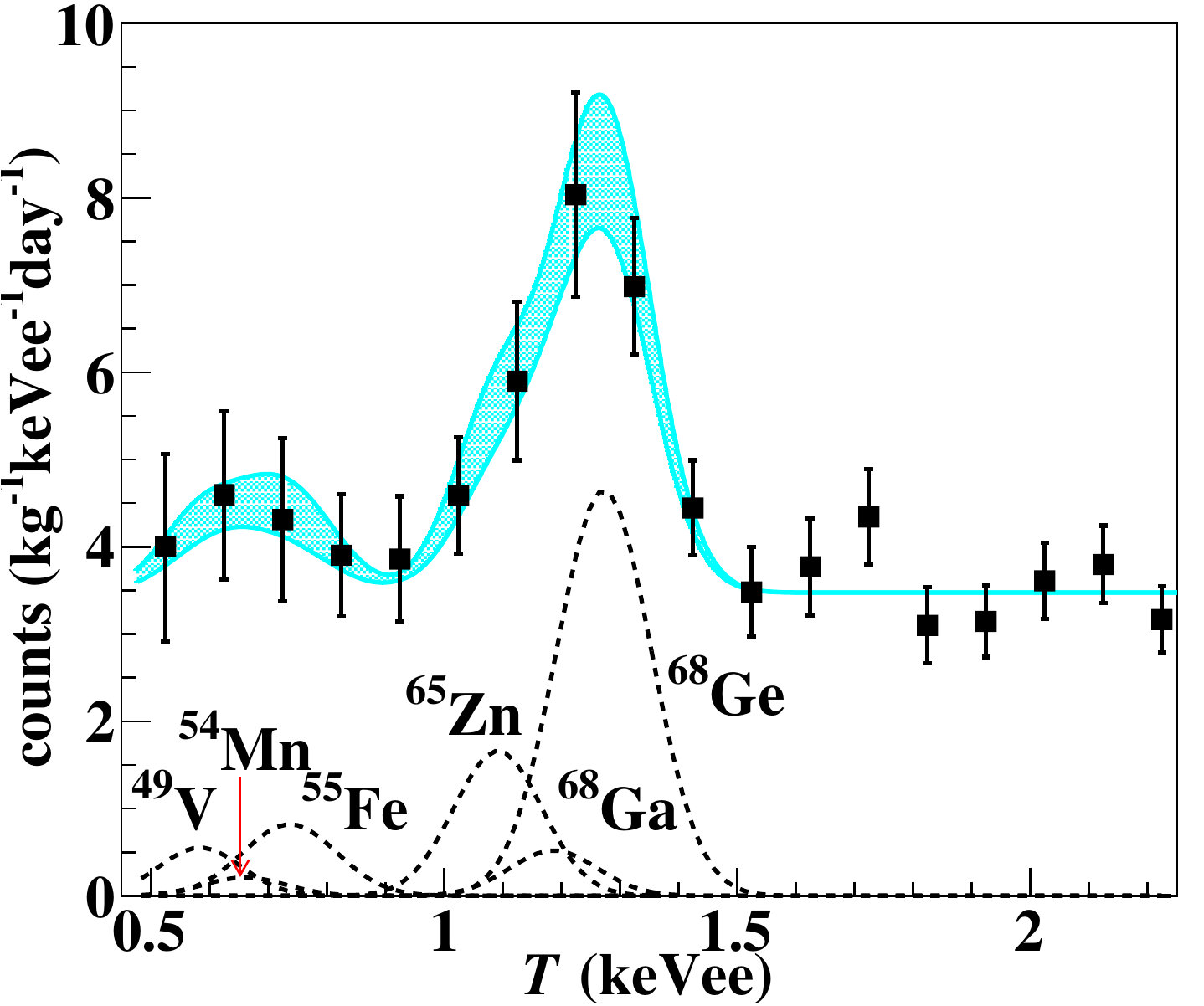}\\
{\bf (c)}\\
\includegraphics[width=7.0cm]{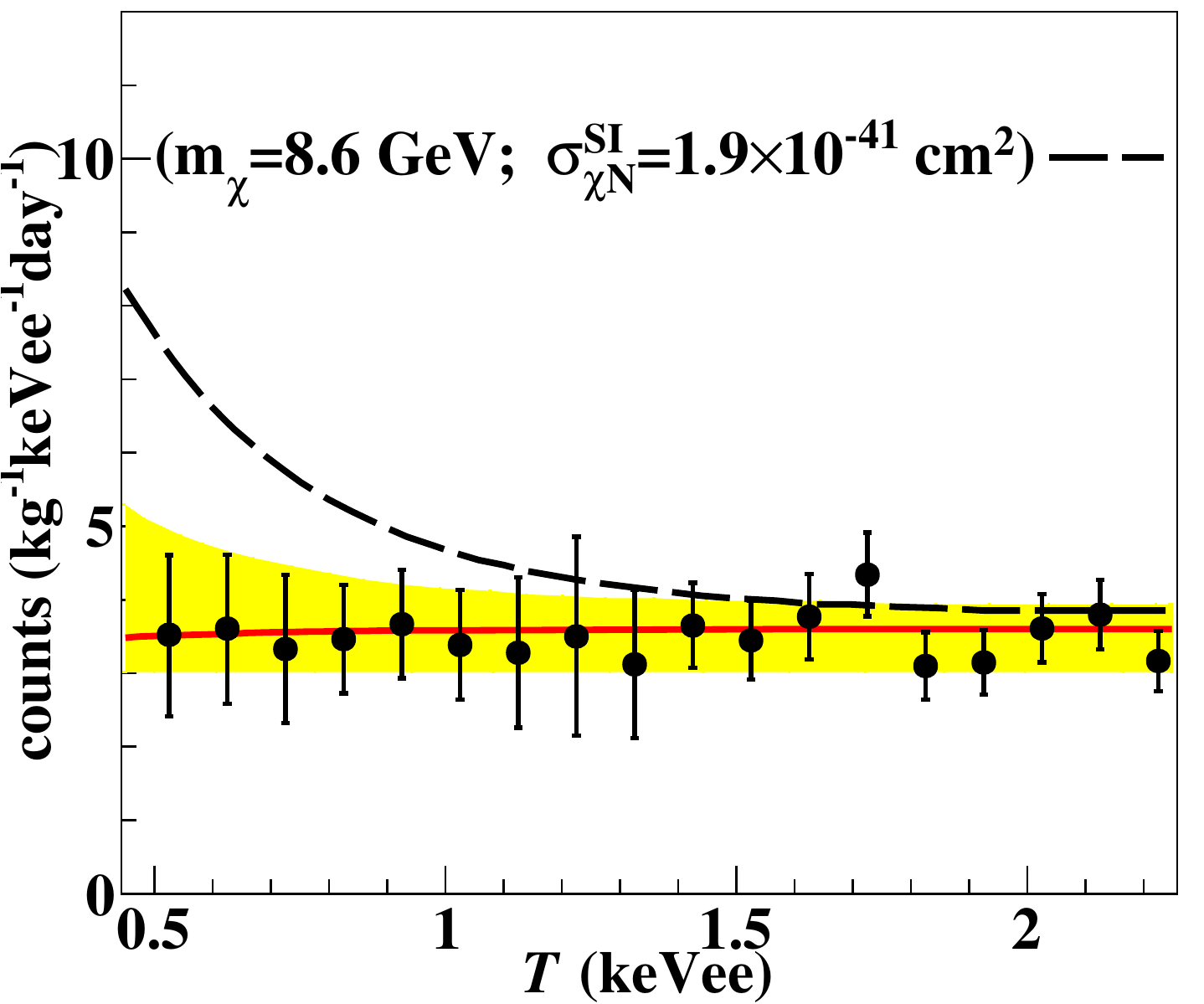}
\caption{\label{fig::cdex1-spectra}
(a)
Background spectra of the CDEX-1 measurement~\cite{cdex}
at their various stages of selection:
basic cuts (TT+Ped+PSD), Anti-Compton (AC) and Bulk (BS) events.
(b)
All events can be accounted for with 
the known background channels $-$ L-shell X-rays and flat background
due to ambient high energy $\gamma$-rays. 
(c)
Residual spectrum with known background channels subtracted.
An excluded $\chi$N recoil spectrum with parameters listed 
is superimposed.
}
\end{center}
\end{figure}


\subsubsection{First Generation CDEX Experiments}

Following schematics of the 
shielding structures and target detectors
depicted in, respectively,
Figures~\ref{fig::ksnlsite}b\&\ref{fig::ge-blndesign},
the first-generation experiments adopted 
the KSNL baseline design~\cite{texonomunu,texonocdm2009} 
of single-element ``1-kg mass scale''
p-PCGe enclosed by NaI(Tl) crystal 
scintillator as anti-Compton detectors,
further surrounded by passive shieldings and
radon purge system. 
Active cosmic-ray vetos are not necessary at this depth.

The pilot CDEX-0 measurement is based on 
the 20~g ULEGe detector array~\cite{texonocdm2009}
at an exposure of 0.784~kg-days
and a threshold of 177~$\rm(eV_{ee})$~\cite{cdex0}.
The CDEX-1 experiment
adopts a p-PCGe detector of mass 1~kg.
The latest results are based on 
an analysis threshold of 475~$\rm{eV_{ee}}$ 
with an exposure of 335.6~kg-days\cite{cdex}.
After suppression of the anomalous surface background events
and measuring their signal efficiencies and background 
leakage factors with calibration data~\cite{bsel2014}, 
all residual events
can be accounted for by known background models.
The spectra are depicted in Figures~\ref{fig::cdex1-spectra}a,b\&c. 
Dark Matter constraints on $\chi$N spin-independent
cross-sections were derived for both data set, 
and are displayed in Figures~\ref{fig::explot}a\&b,
together with other selected benchmark results~\cite{cdmpdg14}. 
In particular, the allowed region from 
the CoGeNT\cite{cogent} experiment 
is probed and excluded with the CDEX-1 results.

Analysis is currently performed on CDEX-1 data set with year-long
exposure.
Annual modulation effects as well as other physics channels 
are being studied. New data is also taken with an upgraded p-PCGe 
with lower threshold.

\subsubsection{Current Efforts and Future Goals}

\begin{figure}
\begin{center}
\includegraphics[width=8.0cm]{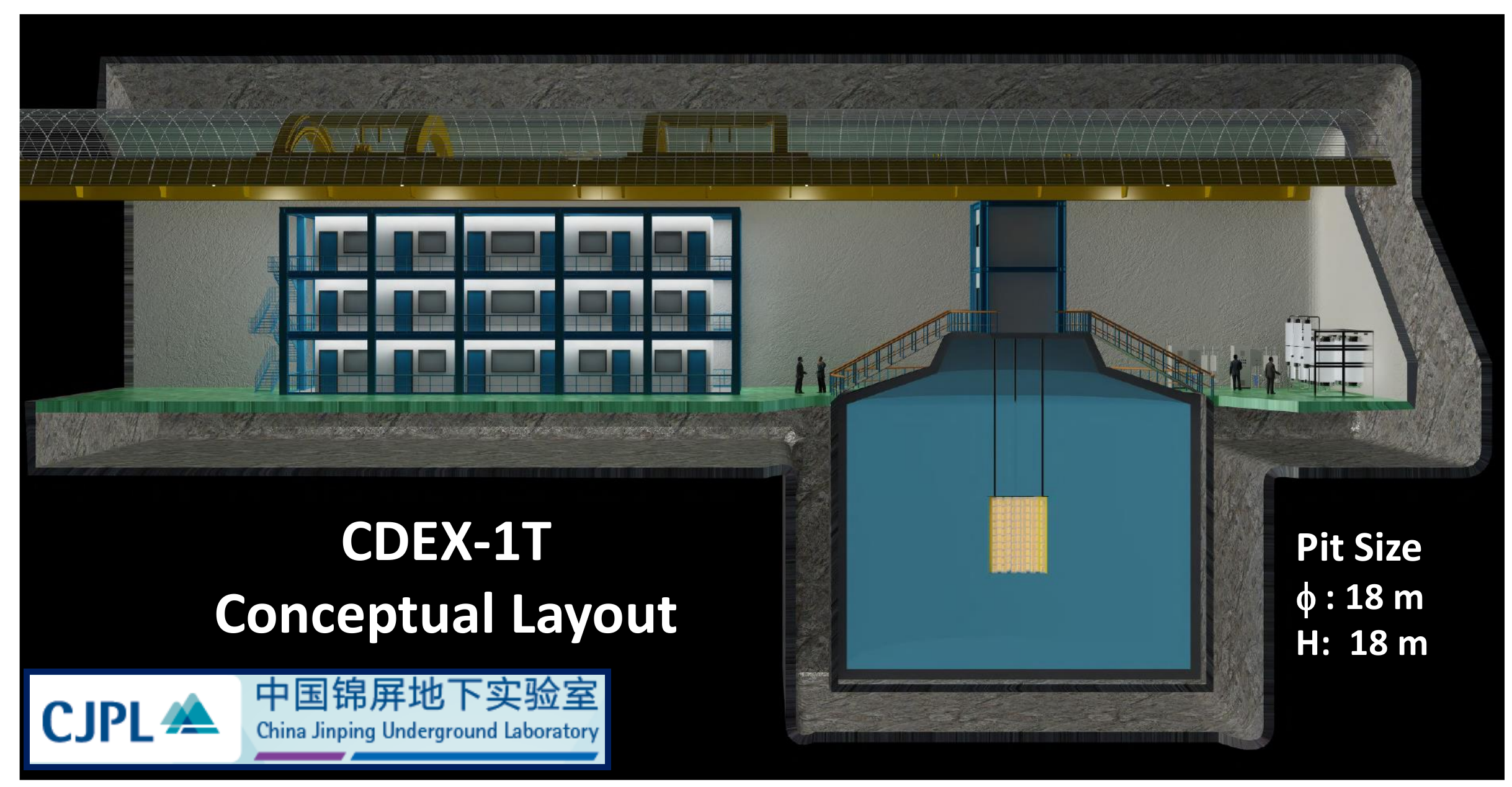}
\caption{\label{fig::cdex1t} 
Conceptual configuration of a future
CDEX-1T experiment at CJPL-II,
showing the pit in which the detector, cryogenic liquid
and water shielding can be placed.
}
\end{center}
\end{figure}

The long-term goal of the CDEX program will be
a ton-scale germanium experiment (CDEX-1T)
at CJPL for the
searches of dark matter and of neutrinoless
double beta decay ($0 \nu \beta \beta$)\cite{0nubb}.
A pit of diameter 18~m and height 18~m is being built
at one of the halls of CJPL-II to house such
an experiment, as illustrated in Figure~\ref{fig::cdex1t}.
The conceptual design is a central region for
germanium detector arrays, surrounded by cryogenic liquid
and/or water shielding.

Towards this ends, the ``CDEX-10'' prototype  
has been constructed with detectors in array structure
having a target mass at the 10-kg range.
This would provide a platform to study
the many issues of scaling up in detector mass and in 
improvement of background and threshold.
The detector array is shielded and cooled by a cryogenic liquid.
Liquid nitrogen is being used, while liquid argon is 
a future option to investigate, which may offer the
additional potential benefits of 
an active shielding as anti-Compton detector.

In addition, various crucial technology acquisition projects
are pursued, which would make a ton-scale germanium experiment 
realistic and feasible. 
These include: 
\begin{enumerate}
\item detector grade germanium crystal growth;
\item germanium detector fabrication;
\item isotopic enrichment of $^{76}$Ge for $0 \nu \beta \beta$;
\item production of electro-formed copper, eventually underground at CJPL.
\end{enumerate}

The first detector fabricated by the CDEX program from
commercial crystal that matches expected performance
is scheduled to be installed at CJPL in 2016. 
It allows control of
assembly materials placed at its vicinity, known to be
the dominant source of radioactive background, while
providing an efficient test platform 
of novel electronics and readout schemes. 
The benchmark would be to perform light WIMP searches
with germanium detectors with ``$0 \nu \beta \beta$-grade''
background control. 
This configuration would provide the
first observation (or stringent upper bounds) 
of the potential cosmogenic tritium contaminations in 
germanium detectors, from which 
the strategies to suppress such background
can be explored.

The projected sensitivity in
$\chi$N spin-independent interactions
for CDEX-1T
is shown in Figure~\ref{fig::explot}a, 
taking a realistic minimal
surface exposure of six months. 

The studies of $0 \nu \beta \beta$ can 
address the fundamental question
on whether the neutrinos are Dirac or Majorana
particles~\cite{0nubb}.
The current generation of
$^{76}$Ge-$0 \nu \beta \beta$ experiments~\cite{gerda+mjd}
are among those with leading sensitivities in
the pursuit.
The objective of a possible 
``CDEX-1T@CJPL-II'' experiment
in $0 \nu \beta \beta$ will be 
to achieve sensitivities sufficient 
completely cover 
the inverted neutrino mass hierarchy~\cite{pdg-nuosc}.
An international network is emerging
towards the formation of a proto-collaboration
to pursue this challenging goal.

Such visions stand on 
several important merits and 
deserve serious considerations.
The overburden at CJPL is among the deepest in
the world, essential for the unprecedented 
background control requirements for such a 
project. Being a new laboratory, there are ample space 
for possible Ge-crystal growth and 
detector fabrication and copper production,
in addition to the pit for 
the main detector and shielding in Figure~\ref{fig::cdex1t}. 
Furthermore, the crucial aspect of this project
is the necessity of 
delivering industrial standard practices 
and control during the mass production 
of detector hardware.
This requirement matches well to the profile
of the CDEX-THU group, being closely 
associated with an industry~\cite{nuctech} 
which has strong experience in
the construction and deployment of
large radiation detector systems 
for international clients.

\section{Prospects}

The TEXONO Collaboration 
was launched from an
operating system without 
previous traditions 
and infra-structures and experience
of running its own particle physics
experiments.
With almost two decades of dedications
and persevering efforts, 
the Collaboration has grown and thrived and emerged into 
having a recognizable presence in the
world stage.
It has contributed to the advances 
and opening of new and innovative avenues in
low energy neutrino physics, light dark matter
searches as well as low threshold germanium detector techniques.
The TEXONO-DNA is propagating in China, India and Turkey
as the alumni members are setting up the 
first-generation research efforts on 
the low-background experimentation 
in these countries.

The observation of neutrino-nucleus coherent scattering
is the top-priority science goal at KSNL.
Scaling that summit would require further advancing
the technologies and techniques
of the germanium detectors.
While the realization of the
underground facility CJPL $-$ and
the implicit investment and commitment $-$ 
is an impressive (and intimidating) feat on its own, 
the science programs are still in their
embryonic and formative stage.
The TEXONO group is at a favorable vantage position
to explore and define and formulate the
future research programs at CJPL
which may bring the Collaboration
to higher grounds.
In particular, 
discussions and studies have been initiated on 
a ton-scale $^{76}$Ge $0 \nu \beta \beta$ 
program at CJPL.

There are grounds to expect that
the future of the TEXONO story,
in spite of $-$ or perhaps because of $-$
not having the detailed landscape charted out yet, 
would be as exciting and rewarding.

\section{Acknowledgments}

The author is whole-heartedly grateful
to the collaborators of both the TEXONO
and CDEX groups, 
the technical and
administrative staff of Academia 
Sinica and the collaborating institutes,
as well as the supporting staff of
the Kuo-Sheng Nuclear Power Station
and the
Yalong River Hydropower Development Company,
for the various invaluable contributions
which ``made these happen''.
Funding support is provided by
Academia Sinica and 
the National Science Foundation,
Ministry of Science and Technology, 
and the National Center of Theoretical Science
in Taiwan,
the National Natural Science Foundation in China,
and the Scientific and Technological Research Council 
in Turkey.


\end{document}